\documentclass{ieeetj}
\usepackage{cite}
\usepackage{amsmath,amssymb,amsfonts}
\usepackage{graphicx,color}
\usepackage{textcomp}
\usepackage[dvipsnames]{xcolor}
\usepackage{hyperref}
\hypersetup{hidelinks=true}
\def\BibTeX{{\rm B\kern-.05em{\sc i\kern-.025em b}\kern-.08em
    T\kern-.1667em\lower.7ex\hbox{E}\kern-.125emX}}
\AtBeginDocument{\definecolor{tmlcncolor}{cmyk}{0.93,0.59,0.15,0.02}\definecolor{NavyBlue}{RGB}{0,86,125}}

\usepackage{etoolbox}

\def\OJlogo{\vspace{-4pt}~}
\def\seclogo{\vspace{10pt}~}

\def\authorrefmark#1{\ensuremath{^{\textbf{#1}}}}

\usepackage{titlesec}       % for \titlespacing*{\section}

\usepackage{balance}        % for \balance at the end of the document
\usepackage{bm}			% for \bm ... bold math 
\usepackage[per-mode=symbol]{siunitx}    		% for \SI{}{} units
\DeclareSIUnit{\dBm}{dBm}	% SI unit "dBm"
\usepackage[
    indexonlyfirst, % only index first use
]{glossaries}
\makenoidxglossaries%
\newacronym{2d}{2D}{two-dimensional}
\newacronym{3d}{3D}{three-dimensional}
\newacronym{ac}{AC}{alternating current}
\newacronym{am}{AM}{amplitude modulation}
\newacronym{aoa}{AOA}{angle-of-arrival}
\newacronym{afe}{AFE}{analog front end}
\newacronym{aod}{AOD}{angle-of-departure}
\newacronym{asic}{ASIC}{application specific integrated circuit}
\newacronym{ambc}{AmBC}{ambient BC}
\newacronym{awgn}{AWGN}{additive white Gaussian noise}
\newacronym{ar}{AR}{augmented reality}
\newacronym{bibc}{BiBC}{bistatic BC}
\newacronym{blue}{BLUE}{best linear unbiased estimator}
\newacronym{ce}{CE}{channel estimation} % needed?
\newacronym{cir}{CIR}{channel impulse response}
\newacronym{cse}{CSE}{channel state estimation}
\newacronym{csi}{CSI}{channel state information}
\newacronym{crlb}{CRLB}{Cram\'er-Rao lower bound}
\newacronym{ct}{CT}{cosine transform} 
\newacronym{cjt}{CJT}{coherent joint transmission}

\newacronym{da}{DA}{data association}
\newacronym{dc}{DC}{direct current}
\newacronym{das}{DAS}{distributed antenna systems}
\newacronym{dli}{DLI}{direct link interference}
\newacronym{dl}{DL}{downlink}
\newacronym{ul}{UL}{uplink}
\newacronym{dm}{DM}{diffuse multipath}
\newacronym{dmc}{DMC}{diffuse multipath component}
\newacronym{dmimo}{D-MIMO}{distributed MIMO}
\newacronym{dtft}{DTFT}{discrete-time Fourier transform}
\newacronym{doa}{DOA}{direction-of-arrival}
\newacronym{dut}{DUT}{device under test}
\newacronym{ec}{EC}{European Commission}
\newacronym{eirp}{EIRP}{equivalent isotropically radiated power}
\newacronym{en}{EN}{energy neutral}
\newacronym{end}{END}{energy neutral device}
\newacronym{epu}{EPU}{edge processing unit}
\newacronym{epc}{EPC}{electronic product code}
\newacronym{esd}{ESD}{electrostatic discharge}
\newacronym{erp}{ERP}{equivalent radiated power}
\newacronym{etsi}{ETSI}{European Telecommunications Standards Institute}
\newacronym{evd}{EVD}{eigenvalue decomposition}
\newacronym{fa}{FA}{federation anchor}
\newacronym{fem}{FEM}{finite element analysis}
\newacronym{fdd}{FDD}{frequency-division duplexing}
\newacronym{fir}{FIR}{finite impulse response}
\newacronym{fss}{FSS}{frequency selective surface}
\newacronym{glrt}{GLRT}{generalized log-likelihood ratio test}
\newacronym{gsm}{GSM}{Global System for Mobile Communications}  % 
\newacronym{eh}{EH}{energy harvesting}
\newacronym{ic}{IC}{integrated circuit}
\newacronym{id}{ID}{information decoding}
\newacronym{iid}{i.i.d.}{independent and identically distributed}
\newacronym{iot}{IoT}{Internet of Things}
\newacronym{ism}{ISM}{industrial, scientific and medical}
\newacronym{jfet}{JFET}{junction field effect transistor}
\newacronym{lis}{LIS}{large intelligent surface}
\newacronym{lmmse}{LMMSE}{linear MMSE}
\newacronym{los}{LoS}{line-of-sight}
\newacronym{ls}{LS}{least-squares}
\newacronym{lti}{LTI}{linear time-invariant}
\newacronym{mc}{MC}{Monte Carlo}
\newacronym{mimo}{MIMO}{multiple-input multiple-output}
\newacronym{miso}{MISO}{multiple-input single-output}
\newacronym{ml}{ML}{maximum likelihood}
\newacronym{mmwave}{mmWave}{millimeter wave}
\newacronym{mmse}{MMSE}{minimum mean square error}
\newacronym{mobc}{MoBC}{monostatic BC}
\newacronym{mosfet}{MOSFET}{metal-oxide semiconductor field effect transistor}
\newacronym{mn}{MN}{matching network}
\newacronym{mpc}{MPC}{multipath component}
\newacronym{music}{MUSIC}{MUltiple SIgnal Classification}
\newacronym{mrt}{MRT}{maximum ratio transmission}
\newacronym{mppt}{MPPT}{maximum power point tracker}
\newacronym{nfc}{NFC}{near-field communication}
\newacronym{nlos}{NLoS}{non-LoS}
\newacronym{olos}{OLoS}{obstructed LoS}
\newacronym{nmos}{nMOS}{n-channel metal-oxide semiconductor}
\newacronym{ofdm}{OFDM}{orthogonal frequency-division multiplexing}
\newacronym{ook}{OOK}{on–off keying}
\newacronym{ota}{OTA}{over-the-air}
\newacronym{nb}{NB}{narrowband}
\newacronym{nr}{NR}{new radio}
\newacronym{pc}{PC}{pilot count}
\newacronym{pce}{PCE}{power conversion efficiency}
\newacronym{pf}{PF}{particle filter}
\newacronym{pg}{PG}{path gain}
\newacronym{pl}{PL}{path loss}
\newacronym{pwm}{PWM}{pulse-width modulation}
\newacronym{rcs}{RCS}{radar cross section}
\newacronym{rf}{RF}{radio frequency}
\newacronym{rfid}{RFID}{radio frequency identification}
\newacronym{ris}{RIS}{reflective intelligent surface}%reconfigurable?
\newacronym{rms}{RMS}{root mean square}
\newacronym{rmse}{RMSE}{root mean square error}
\newacronym{rrmse}{RRMSE}{relative root mean square error}
\newacronym{rss}{RSS}{received signal strength}
\newacronym{rssi}{RSSI}{received signal strength indicator}
\newacronym{rw}{RW}{RadioWeave}
\newacronym{rv}{RV}{random variable}
\newacronym{s-parameter}{S-parameter}{scattering parameter}
\newacronym{sa}{SA}{synchronization anchor}
\newacronym{sar}{SAR}{specific absorption rate}
\newacronym{sbl}{SBL}{sparse Bayesian learning}
\newacronym{sdn}{SDN}{software-defined network}
\newacronym{srd}{SRD}{short-range device}
\newacronym{sinr}{SINR}{signal-to-interference-plus-noise ratio}
\newacronym{sir}{SIR}{signal-to-interference ratio}
\newacronym{siso}{SISO}{single-input single-output}
\newacronym{simo}{SIMO}{single-input multiple-output}
\newacronym{slam}{SLAM}{simultaneous localization and mapping}
\newacronym{snr}{SNR}{signal-to-noise ratio}
\newacronym{smc}{SMC}{specular multipath component}
\newacronym{svd}{SVD}{singular value decomposition}
\newacronym{si}{SI}{self interference}
\newacronym{sw}{SW}{spherical wavefront}
\newacronym{tdd}{TDD}{time division multiplexing}
\newacronym{tdoa}{TDOA}{time-difference-of-arrival}
\newacronym{toa}{TOA}{time-of-arrival}
\newacronym{tosm}{TOSM}{through–open–short–match}
\newacronym{ue}{UE}{user equipment}
\newacronym{uhf}{UHF}{ultra high frequency}
\newacronym{ula}{ULA}{uniform linear array}
\newacronym{upa}{UPA}{uniform planar array}
\newacronym{ura}{URA}{uniform rectangular array}
\newacronym{uv}{UV}{unmanned vehicle}
\newacronym{vna}{VNA}{vector network analyzer}
\newacronym{wb}{WB}{wideband}
\newacronym{wpt}{WPT}{wireless power transfer}
\newacronym{wrt}{w.r.t.}{with respect to}
\newacronym{xets}{XETS}{cross exponentially tapered slot}
\newacronym{zf}{ZF}{zero-forcing}
\newacronym{csp}{CSP}{contact service point}
\newacronym{ecsp}{ECSP}{edge computing service point}
\newacronym{cmos}{CMOS}{complementary metal oxide semiconductor}
\newacronym{ask}{ASK}{amplitude shift keying}
\newacronym{fdma}{FDMA}{frequency division multiple access}
\newacronym{fsk}{FSK}{frequency shift keying}
\newacronym{ber}{BER}{bit error rate}
\newacronym{pdf}{PDF}{probability density function}
\newacronym{psk}{PSK}{phase shift keying}
\newacronym{qam}{QAM}{quadrature amplitude modulation}
\newacronym{cw}{CW}{continuous wave}
\newacronym{papr}{PAPR}{peak-to-average power ratio}
\newacronym{bc}{BC}{backscatter communication}
\newacronym{ca}{CE}{carrier emitter}
\newacronym{bd}{BD}{backscatter device}

\newacronym{uwb}{UWB}{ultra-wideband}
\newacronym{bgmm}{BGMM}{Bayesian Gaussian mixture model}
\newacronym{gmm}{GMM}{Gaussian mixture model}
\newacronym{knn}{kNN}{k-Nearest-Neighbors}
\newacronym{ft}{FT}{Fourier transform}
\newacronym{dft}{DFT}{discrete Fourier transform}

\newacronym{adc}{ADC}{analog-to-digital converter}
\newacronym{np}{NP}{Neyman-Pearson}
\newacronym{pana}{PanA}{Panel A}
\newacronym{panb}{PanB}{Panel B}
\newacronym{p1}{P1}{Phase 1}
\newacronym{p2}{P2}{Phase 2}
\newacronym{pw}{PW}{planar wavefront}
\newacronym{i.i.d.}{i.i.d.}{independent and identically distributed}
\newacronym{pcsi}{PCSI}{perfect channel state information}
\newacronym{mrfh}{MRFH}{main RF harvester}
\newacronym{arfh}{ARFH}{auxiliary RF harvester}
\newacronym{mcu}{MCU}{microcontroller unit}

\newacronym{fim}{FIM}{Fisher information matrix}
\newacronym{efim}{EFIM}{equivalent FIM}
\newacronym{mva}{MVA}{master virtual anchor}
\newacronym{pa}{PA}{physical anchor}
\newacronym{pcrlb}{PCRLB}{posterior CRLB}
\newacronym{peb}{PEB}{positioning error bound}
\newacronym{ceb}{CEB}{clock error bound}
\newacronym{meb}{MEB}{mapping error bound}
\newacronym{pmva}{PMVA}{potential MVA}
\newacronym{va}{VA}{virtual anchor}
\newacronym{vm}{VM}{virtual mobile}

\usepackage{fontawesome}		% for icons (incl. checkmarks)
\usepackage{array, booktabs, xltabular}
\usepackage{colortbl}       % for \cellcolor
\usepackage{multirow}
\let\labelindent\relax  % if the template does not like the {enumitem} package
\usepackage{enumitem}   % to use the leftmargin property (incompatible with beamer)

\usepackage{textcomp}       % for \texttrademark

\usepackage{url}
\newcommand{\icon}[3]{\makebox(#2, #2){\textcolor{#3}{\csname fa#1\endcsname}}}	%icon shortcut
\usepackage{caption}
\usepackage[labelformat=simple]{subcaption}
\usepackage{placeins}
\usepackage{tikzscale}		% scale tikz plot
\usepackage{tcolorbox}

\usepackage{tikz}
\usepackage{ifthen}

\usepackage[framemethod=TikZ]{mdframed}	% for math boxes with 

\usepackage{pgfplots}
  \tcbset{shield externalize} %there can be problems with tikz externalize and tcolorbox
  \pgfplotsset{compat=newest}
  \usetikzlibrary{plotmarks}
  \usetikzlibrary{arrows.meta}
  
  \usetikzlibrary{arrows}
  \usetikzlibrary{mindmap,trees}%,backgrounds}
  \usetikzlibrary{positioning,spy}
  \usetikzlibrary{calc,fadings,decorations.pathreplacing}
  \usetikzlibrary{decorations.pathmorphing,decorations.text}
  \usetikzlibrary{datavisualization.polar}      % for polar plots
  \usepackage{pgfplotstable}
  \usepgfplotslibrary{polar}
  \usetikzlibrary{patterns}

  \usetikzlibrary{plotmarks}
  \usepgfplotslibrary{patchplots}
  \usepackage{grffile}
  \usepackage{amsmath}
  \usepackage{tikzscale}		% scale tikz plot
  \usepackage{tikz-3dplot} 
  \usetikzlibrary{positioning}
  \usetikzlibrary{arrows}
  \usetikzlibrary{fit,backgrounds}  % for the fit of boxes

\pgfdeclarelayer{behindBackground}
\pgfdeclarelayer{background}
\pgfdeclarelayer{foreground}
\pgfsetlayers{behindBackground,background,main,foreground}   %% some additional layers for demo

\tikzset{%
  >=latex,
  inner sep=0pt,%
  outer sep=2pt,%
  mark coordinate/.style={inner sep=0pt,outer sep=0pt,minimum size=3pt,
  fill=black,circle}%
}
\newcommand{\externalizeFigures}{false}
\ifthenelse{\equal{\externalizeFigures}{true}}
{
  \usepackage{pgfplots}
  \pgfplotsset{compat=newest}
  \usepgfplotslibrary{external}
  \tikzexternalize[prefix=tikz/]
  \usepackage{shellesc}
}
{}
\newcommand{\circleblue}[1]{ \tikz[baseline=(char.base)]{
            \node[shape=circle,inner sep=1pt,fill=IEEEblue, text=white] (char) {#1};}}

\newlength{\plotWidth}		% length for scaling the plot
\newlength\figureheight
\newlength\figurewidth

\newcommand{\trajectoryEstimateLW}{1pt}           
\newcommand{\trajectoryEstimateCOL}{IEEEred}          
\newcommand{\trajectoryLW}{1.0pt}     % line width of the trajectories

\newcommand{\herm}{\mathsf{H}}

\newcommand{\trp}{\mathsf{T}}

\DeclareMathOperator{\vect}{vec}

\DeclareMathOperator*{\argmax}{arg\,max}
\DeclareMathOperator{\diag}{diag}

\newcommand{\realset}[2]{ \mathbb{R}^{#1 \times #2}  }

\newcommand{\complexset}[2]{ \mathbb{C}^{#1 \times #2}  }

\DeclareMathOperator{\tr}{tr}
\DeclareMathOperator{\rank}{rank}

\usepackage{amsthm}
\newtheorem{proposition}{Proposition}

\newcommand{\customlineref}[1]{
    \tikz[baseline={([yshift=-.5ex]current bounding box.center)}, inner sep=0, outer sep=0,xshift =0pt]{
        \hspace*{-0.2em} % Adjust this value to reduce space on the left
        \draw[#1] (0,0) -- (0.6cm,0);
    }\hspace*{-0.4em}
}

\usepackage{algorithm2e}		% for \begin{algorithm} ... \end{algorithm}
\RestyleAlgo{ruled}			% makes rules around the algorithms
\SetKw{KwBy}{by}			% 
\SetKwComment{Comment}{\% }{}	% for \Comment*[r]{ <text> }

\usepackage[hang,flushmargin]{footmisc}
\makeatletter
\newcommand{\algorithmfootnote}[2][\footnotesize]{%
  \let\old@algocf@finish\@algocf@finish% Store algorithm finish macro
  \def\@algocf@finish{\old@algocf@finish% Update finish macro to insert "footnote"
    \leavevmode\rlap{\begin{minipage}{\linewidth}
    #1#2
    \end{minipage}}%
    \vspace{-0.3cm} % [B] hack :')
  }%
}

\DeclareMathOperator*{\SNR}{SNR}

\definecolor{RDlightgreen}{RGB}{141 192 69}
\definecolor{RDgreen}{rgb}{0.3647, 0.4275, 0.2667}
\definecolor{RDdarkgreen}{rgb}{0.2196, 0.2196, 0.2196}
\definecolor{RDmaroon}{rgb}{.522,.22,.353} %

\definecolor{IEEEblue}{RGB}{0 98 155}
\definecolor{IEEElightblue}{RGB}{0 181 226}
\definecolor{IEEEturquoise}{RGB}{0 156 166}
\definecolor{IEEEred}{RGB}{186 12 47}
\definecolor{IEEEgreen}{RGB}{0 132 61}
\definecolor{IEEElightgreen}{RGB}{120 190 32}
\definecolor{IEEEorange}{RGB}{225 163 0}
\definecolor{IEEEyellow}{RGB}{255 209 0}
\definecolor{IEEEviolett}{RGB}{152 29 151}
\definecolor{IEEEdarkmaroon}{RGB}{134 31 65}

\definecolor{myblue}{rgb}{0.00000,0.60580,0.87650}  % Fused CSI

\newcolumntype{C}{@{\hskip 0.075cm}c@{\hskip 0.075cm}}

\newcommand{\eye}[1]{\mathbf{I}_{\scriptscriptstyle #1}}                % identity-matrix of size (#1 x #1)
\newcommand{\diff}[0]{\mathrm{d}}       

\newcommand{\fc}{f_{\text{\tiny c}}}                                % carrier frequency
\newcommand{\lightspeed}{\mathsf{c}}                                % vacuum speed of light

\newcommand{\Nrw}{J}                                                % number of anchors
\newcommand{\Nfrequency}[0]{ N_{\scriptscriptstyle\!f} }             % number of frequency bins
\newcommand{\nfrequency}[0]{ n_{\scriptscriptstyle\!f} }             % frequency bin index
\newcommand{\Nantennas}[0]{ M }                                     % number of antennas
\newcommand{\nantenna}[0]{ m }                                      % antenna index
\newcommand{\Nwalls}[0]{ S }                                        % number of walls
\newcommand{\Ncomponents}[0]{ K }                          % number of components (LoS+SMCs) 
\newcommand{\ncomponent}[0]{ \kappa }                               % current component index
\newcommand{\stateSpace}[0]{\mathcal{S}_{\scriptscriptstyle\!\bm{\theta}}}                            % the state space
\newcommand{\Nparticles}[0]{ N_{\text{\tiny p}} }            % number of particles

\newcommand{\setWalls}[0]{ \mathcal{S} }                            % set of Walls
\newcommand{\setSDB}[0]{ \mathcal{D} }                              % set of Single AND Double Bounce Paths
\newcommand{\setSDBt}[0]{ \widetilde{\mathcal{D}} }                 % combined set of LoS, Single Bounce Paths and Double Bounce Paths
\newcommand{\setSB}[0]{ \mathcal{D}_{\text{S}} }                    % set of Single Bounce Paths
\newcommand{\setDB}[0]{ \mathcal{D}_{\text{D}} }                    % set of Double Bounce Paths

\newcommand{\MVApair}[2]{(\!#1,#2\!)}                               % pair of MVAs

\newcommand{\Pjss}[3]{\bm{P}_{\scriptscriptstyle #1}^{\MVApair{#2}{#3}}}                           % array layout physical/mirror anchor

\newcommand{\pmva}[1]{\bm{p}^{\text{\tiny mva}}_{\scriptscriptstyle #1}}               % position of MVA #1
\newcommand{\hva}[0]{{h}_{\text{\tiny va}}}                         % MVA -> VA transformation
\newcommand{\origin}[0]{\bm{0}}                                     % origin of the coordinate system
\newcommand{\rotM}[0]{\bm{M}_{\scriptscriptstyle j}}                                       % rotation Matrix
\newcommand{\posAnchor}[3]{ {\bm{p}_{\text{\tiny c}\scriptscriptstyle,#1}^{\scriptscriptstyle\MVApair{#2}{#3}}} }     % position of the RW center of gravity in global coordinates
\newcommand{\house}[1]{ \bm{\mathcal{H}}_{\scriptscriptstyle #1} }   % householder matrix for MVA #1

\newcommand{\PiPar}[2]{\bm{\Pi}_{\scriptscriptstyle \MVApair{#1}{#2}}^{\parallel}}  % parallel polarization projector on surface #1
\newcommand{\PiPerp}[2]{\bm{\Pi}_{\scriptscriptstyle \MVApair{#1}{#2}}^{\perp}}  % orthogonal polarization projector on surface #1
\newcommand{\observation}[2]{\bm{z}_{\scriptscriptstyle #1}^{\scriptscriptstyle(#2)}}   % observation of anchor #2 at time step #1
\newcommand{\observationMatrix}[1]{\bm{Z}_{\scriptscriptstyle #1}}                      % observation of ALL anchors at time step #1
\newcommand{\noise}[2]{\bm{w}_{\scriptscriptstyle #1}^{\scriptscriptstyle(#2)}}         % noise of anchor #2 at time step #1
\newcommand{\dictionary}[1]{\bm{\Psi}_{\scriptscriptstyle #1}}                          % dictionary matrix of #1-th anchor
\newcommand{\dictEntry}[2]{\bm{\psi}_{\scriptscriptstyle j}^{\scriptscriptstyle \MVApair{#1}{#2}}}         % dictionary component of MVA pair (#1,#2)
\newcommand{\PMVAposStacked}[1]{\overline{\bm{p}}^{\text{\tiny mva}}_{\scriptscriptstyle \,}}                   % stacked positions of ALL MVAs at time #1 
\newcommand{\state}[1]{\bm{x}_{\scriptscriptstyle #1}}                                  % agent state at time step #1
\newcommand{\stateEstimate}[1]{\hat{\bm{x}}_{\scriptscriptstyle #1}}                                  % agent state estimate at time step #1
\newcommand{\covarianceEstimate}[1]{\hat{\bm{P}}_{\scriptscriptstyle#1}}                % state covariance estimate at time instance #1
\newcommand{\pos}[1]{ \bm{p}_{\scriptscriptstyle #1} }                                  % physical agent position at time step #1
\newcommand{\posEstimate}[1]{ \hat{\bm{p}}_{\scriptscriptstyle #1} }                    % estimated agent position at time step #1
\newcommand{\alphavec}[1]{\bm{\alpha}_{\scriptscriptstyle #1}}                       % complex-valued amplitude vector at time #1 
\newcommand{\mualpha}[1]{\bm{\mu}_{\scriptscriptstyle %\alpha,
#1}}                       % complex-valued amplitude vector at time #1 
\newcommand{\sigmaHat}[1]{\widehat{\sigma_{\scriptscriptstyle #1}^{\scriptscriptstyle2}}} % estimate of the circular AWGN variance at anchor #1
\newcommand{\sourceCov}[1]{\bm{P}_{\scriptscriptstyle #1}}                       % source covariance matrix 
\newcommand{\Phat}[2]{\widehat{\bm{P}}_{\scriptscriptstyle #1,#2}}                  % estimate of the source covariance matrix at time #1 of anchor #2
\newcommand{\signalCov}[1]{\bm{R}_{\scriptscriptstyle #1}}                       % spatiotemporal signal covariance matrix 
\newcommand{\Rhate}[2]{\widehat{\bm{R}}_{\scriptscriptstyle #1,#2}^{\text{\scriptsize e}}} % empirical covariance matrix at time #1 of anchor #2
\newcommand{\Rhat}[2]{\widehat{\bm{R}}_{\scriptscriptstyle #1,#2}} % STO covariance matrix at time #1 of anchor #2

\newcommand{\alphavecHat}[1]{\widehat{\bm{\alpha}}_{\scriptscriptstyle #1}}                       % complex-valued amplitude vector estimate at time #1 
\newcommand{\pathgain}[0]{G}                             % pathgain symbol

\newcommand{\particle}[2]{\bm{\theta}_{\scriptscriptstyle #1}^{\scriptscriptstyle(#2)}}     % particle #2 at time instance #1
\newcommand{\weight}[2]{w_{\scriptscriptstyle #1}^{{\scriptscriptstyle(#2)}}}          % weight of particle #2 at time instance #1
\newcommand{\jacobian}{ \boldsymbol{J} }
\newcommand{\jacobMVAsb}[1]{ \jacobian_{\scriptscriptstyle n,j}^{\scriptscriptstyle \text{\tiny sb},s}}   % noise-free signal model at time n, anchor j for component (#3,#4)

\newcommand{\velocity}[1]{ \bm{v}_{\scriptscriptstyle #1} }                            % velocity vector of EN device at time step #1
\newcommand{\pwk}[0]{ \bm{p}^\text{\tiny{w}}_{\scriptscriptstyle s} }                  % arbitrary point on the k-th wall
\newcommand{\nw}[0]{ {\bm{n}^\text{\tiny{w}}_{\scriptscriptstyle s}} }                   % normal vector on the s-th wall
\newcommand{\Ptemplate}[0]{\bm{P}_{\text{t}}}                       % sensor layout template aperture

\newcommand{\posantenna}[1]{ {\bm{p}^{\text{\tiny a}}_{\scriptscriptstyle #1}} }       % position of an antenna w.r.t. the center of gravity of the RW
\newcommand{\transitionmatrix}{{\bm{\Phi}}}                         % state transition matrix
\newcommand{\processNoiseCov}{{\bm{Q}}}                             % process noise covariance matrix
\newcommand{\processNoise}[1]{{\bm{\omega}_{\scriptscriptstyle #1}}}% process noise vector at time #1
\newcommand{\channel}{\bm{h}}                                       % general channel vector
\newcommand{\hestimate}[3]{\widetilde{\bm{h}}_{\scriptscriptstyle #1,#2,#3}} % f-domain sum channel estimate at antenna #1, time #2, anchor #3
\newcommand{\hcomponent}[5]{\bm{h}_{\scriptscriptstyle #1,#2,#3}^{\MVApair{#4}{#5}}} % f-domain channel component at antenna #1, time #2, anchor #3, mva-pair #4-#5
\newcommand{\hpredict}[1]{{\channel}_{\scriptscriptstyle #1}^{\scriptscriptstyle \star}}                        % channel prediction at time #1
\newcommand{\hmeas}[1]{\widetilde{\channel}_{\scriptscriptstyle #1}}                        % channel prediction at time #1
\newcommand{\hfused}{\hat{\channel}_{\text{\tiny f}}}               % fused channel
\newcommand{\onematrix}[2]{\bm{1}_{\scriptscriptstyle{#1\times#2}}}        % matrix of ones

\newcommand{\nsubarray}{j}                         % anchor index

\newcommand{\possubarray}[1]{ {\bm{p}^{\text{\tiny c}}_{#1}} }     % position of the RW center of gravity in global coordinates
\newcommand{\posSMC}[2]{ {\bm{p}_{\scriptscriptstyle #2,\text{\tiny va}}^{\scriptscriptstyle(#1)} }}   % mirror anchor center of gravity position

\renewcommand{\state}[1]{\bm{\theta}_{\scriptscriptstyle\!#1}}                                  % joint state at time step #1
\newcommand{\agentState}[1]{\bm{x}_{\scriptscriptstyle #1}}                             % agent state at time step #1
\renewcommand{\stateEstimate}[1]{\hat{\bm{\theta}}_{\scriptscriptstyle #1}}                    % joint state estimate at time step #1
\renewcommand{\PiPar}[1]{\bm{\Pi}_{\scriptscriptstyle #1}^{\parallel}}  % parallel projector on the dictionary of anchor #1
\renewcommand{\PiPerp}[1]{\bm{\Pi}_{\scriptscriptstyle #1}^{\perp}}  % perpendicular projector on the dictionary of anchor #1

\newcommand{\htrue}[1]{\channel_{\scriptscriptstyle #1}}                            % true CSI
\renewcommand{\hfused}[1]{{\channel}^{\text{\tiny f}}_{\scriptscriptstyle #1}}      % fused CSI
\renewcommand{\hpredict}[1]{{\channel}^{\text{\tiny p}}_{\scriptscriptstyle #1}}    % predicted CSI

\newcommand{\shortPaperVersion}{true} % set true for page-limited publication or false for long (e.g., arXiv) version

\titlespacing*{\section}{0pt}{1ex}{0ex}
\titlespacing*{\subsection}{0pt}{1ex}{0ex}
\titlespacing*{\subsubsection}{0pt}{0.1ex}{0.5ex}

\begin{document}
\setlength{\intextsep}{15pt}     % Space above and below in-text floats
\setlength{\textfloatsep}{15pt}   % Space between floats and text
\setlength{\floatsep}{15pt}       % Space between floats on a page
\colorlet{IEEEblue}{NavyBlue}			% Overwrite IEEEblue with the Journal blue

%\receiveddate{~}
%\reviseddate{~}
%\accepteddate{~}
%\publisheddate{~}
%\currentdate{~}
%\doiinfo{~}

\markboth{}{Author {et al.}}

\title{Geometry-Based Channel\\Estimation, Prediction, and Fusion
}

%\author{Benjamin J.\,B. Deutschmann\authorrefmark{1}, Student Member, IEEE, % 1st author, 1st affiliations
%Erik Leitinger\authorrefmark{1}, Member, IEEE, and % 2nd 
%Klaus Witrisal\authorrefmark{1}, Member, IEEE}     % last author, affiliation

\author{Benjamin J.\,B. Deutschmann\authorrefmark{1}, % 1st author, 1st affiliations
Erik Leitinger\authorrefmark{1}, and % 2nd 
Klaus Witrisal\authorrefmark{1}}     % last author, affiliation

\affil{Graz University of Technology, Graz, Austria}
\corresp{Corresponding author: Benjamin J.\,B. Deutschmann (email: benjamin.deutschmann@tugraz.at).}
%\authornote{The project has received funding from the EU's Horizon Europe research and innovation program under grant agreement No~101192113.}
\authornote{The AMBIENT-6G project has received funding from the Smart Networks and Services Joint Undertaking (SNS JU) under the European Union's Horizon Europe research and innovation programme under Grant Agreement No. 101192113.}

\begin{abstract}
Reciprocity-based beamforming---most commonly employed in time-division duplexing---uses noisy, estimated (i.e., measured) channel state information (CSI) acquired on the uplink.
While computationally efficient, reciprocity-based beamforming suffers severe losses under (i) low signal-to-noise ratio (SNR) and (ii) user mobility because it ignores the underlying physics of the radio channel beyond its reciprocity. 
Based on a physics-driven geometry-based channel model, we propose a method that jointly infers the mobile user's position and environment map on the uplink. 
It then leverages the estimated user position and environment map to predict CSI on the downlink. 
We demonstrate significant efficiency gains under both (i) low SNR and (ii) user mobility on measured data. 
While the user position may allow efficient beamforming in strong line-of-sight (LoS) channels, inferring an environment map allows bypassing obstructed LoS conditions using non-LoS beamforming via multipath components. 
We further propose ``channel fusion,'' a probabilistic (Bayesian) combination of estimated and predicted CSI, which increases the beamforming robustness, particularly when either source of CSI is unreliable. 
Notably, this approach shares similarities with the minimum mean square error (MMSE) channel estimator, with geometry-based prior parameters inferred from the data.
\end{abstract}


\maketitle

\section{INTRODUCTION}\label{sec:introduction}

\IEEEPARstart{C}{hannel} estimates acquired on the \gls{ul} are among the most commonly employed \gls{csi} for \gls{dl} beamforming (i.e., reciprocity-based beamforming) in time-division duplexing~\cite{Flordelis18FDD}.
Channel prediction finds applications in ``extrapolating'' \gls{csi} to time or frequency points where observations have not been made~\cite{Palleit11Prediction}. 
Prominent examples include predicting future \gls{csi} to mitigate \gls{csi} aging under user mobility~\cite{Loeschenbrand23CSIaging}, or predictions from \gls{csi} observed on uplink frequencies to unobserved downlink frequencies in frequency-division duplexing~\cite{Arnold19FDDprediction}.
We use the term ``channel fusion'' for methods that incorporate prior information to improve upon a noisy channel observation, or provide robustness if either the channel estimate or prior information are unreliable.
\begin{figure}%
    \centering%
    \includegraphics[width = \linewidth]{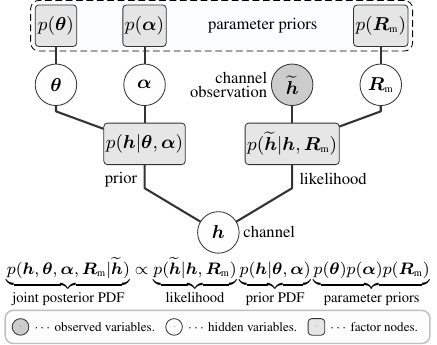}%
    \caption{Factor graph representing channel fusion.}%
    \label{fig:factor-graph}%
\end{figure}%
To appreciate the terminology, consider a parametric multipath channel model~\cite{richter2005estimation}
\begin{align}
    \bm{h} = \bm{\Psi}(\bm{\theta}) \bm{\alpha}
\end{align}
that models $\Ncomponents$ signal sources with amplitudes $\bm{\alpha}\!\in\!\mathbb{C}^{\Ncomponents \times 1}$ propagated over a dictionary\ifthenelse{\equal{\shortPaperVersion}{true}}%
{%      Omit in the short paper
}{\footnotemark}
$\bm{\Psi}\!\in\!\mathbb{C}^{N \times \Ncomponents}$ that is parameterized by a set of geometry-related parameters $\bm{\theta}$. 
With stochastic amplitudes $\bm{\alpha}\!\sim\!p(\bm{\alpha})$ and parameters $\bm{\theta}\!\sim\!p(\bm{\theta})$ the channel prior \gls{pdf} is $p(\bm{h}|\bm{\theta},\bm{\alpha})$. %=\mathcal{CN}(\bm{\Psi}\bm{\mu},\bm{\Psi}\bm{P}\bm{\Psi}^H)$. 
Channel observations $\widetilde{\bm{h}}=\bm{h}+\bm{n}$ are made in additive noise $\bm{n}|\bm{R}_{\text{\tiny m}} \sim \mathcal{CN}(\bm{0},\bm{R}_{\text{\tiny m}})$, hence the observation likelihood is $p(\widetilde{\bm{h}}|\bm{h},\bm{R}_{\text{\tiny m}})$.
The factor graph~\cite{Loeliger04IntroFG} in Fig.\,\ref{fig:factor-graph} represents the joint posterior distribution as the product of its factors.
It can be shown to factorize (up to a proportionality constant) into a product of likelihood, prior, and parameter prior \glspl{pdf} by exploiting the conditional independence structure that the factor graph makes explicit.
We define the terminology: %
\begin{itemize}[leftmargin=15pt]
    \item \textit{Channel estimation}: Estimated \gls{csi} is computed through the maximum
    \begin{align}\label{eq:intro-h-estimated}
        \hat{\bm{h}}_{\text{\tiny m}} = \argmax_{\bm{h}} \big( p(\widetilde{\bm{h}}| \bm{h}) \big) 
        \triangleq
        \widetilde{\bm{h}}
    \end{align}
    of the marginal likelihood $p(\widetilde{\bm{h}}| \bm{h})$ and corresponds to the observed (i.e., measured) \gls{csi} $\widetilde{\bm{h}}$.
    \item \textit{Channel prediction}: Predicted \gls{csi} is computed through the expectation
    \begin{align}\label{eq:intro-h-predicted}
        \hat{\bm{h}}_{\text{\tiny p}} = \mathbb{E} \big( \bm{h} \big)
    \end{align}
    under the marginal prior \gls{pdf} $p(\bm{h})$.
    \item \textit{Channel fusion}: Fused \gls{csi} is computed through the expectation
    \begin{align}\label{eq:intro-h-fused}
        \hat{\bm{h}}_{\text{\tiny f}} = \mathbb{E} \big( \bm{h}|\widetilde{\bm{h}} \big)
    \end{align}
    under the marginal posterior \gls{pdf} $p(\bm{h}|\widetilde{\bm{h}})\propto p(\widetilde{\bm{h}}| \bm{h}) p(\bm{h})$.
\end{itemize}
\ifthenelse{\equal{\shortPaperVersion}{true}}%
{%      Omit in the short paper
}{\footnotetext{The $K$ dictionary entries are elements of the array manifold in e.g. angular, delay, and/or Doppler domains.}}%

\noindent\null%
{%
\setlength{\parindent}{0pt}%
A prominent example of channel fusion is \gls{mmse} channel estimation~\cite{mMIMObook} which often assumes \textit{known} prior parameters, meaning that the model parameters are replaced with constants.
}%
Suppose the parameters cannot be assumed known. 
There are two popular approximate but closed-form methods to treat the parameter prior \glspl{pdf} for obtaining the marginal \glspl{pdf} in~\eqref{eq:intro-h-estimated}\,-\,\eqref{eq:intro-h-fused} that evade a potentially expensive direct marginalization: %from the data: 
Marginalization exploiting conjugate parameter priors~\cite[Sec.\,2.4.2]{Bishop}, or concentration at ML estimates of the parameters computed from the data (i.e., empirical Bayes)~\cite{Larsson07MMSEsparse}.
The literature on methods related to what we term channel fusion can be divided into categories that differ in how the channel prior \gls{pdf} is modeled:%
\begin{enumerate}[leftmargin=15pt]
\item[A.] Model-based approaches: 
     %A.~Model-based approaches 
    \begin{enumerate}[leftmargin=2pt]
        \item[i)] Models based on a %statistical, physics-agnostic/
        geometry-agnostic channel prior $p(\bm{h})$ possibly inferring parameter priors online from the data~\cite{Larsson07MMSEsparse,Ozdogan19MMSE,Kashyap2017Kalman,Kim21KFvsMachineLearning,Palleit11Prediction}.
        Prominent members of this category are works that first predict \gls{csi} (to future time steps) using an autoregressive model and then fuse it with estimated \gls{csi} using a Kalman filter directly in the space of the channel $\bm{h}$ rather than the parameters $\bm{\theta}$~\cite{Kashyap2017Kalman,Kim21KFvsMachineLearning}.
        \item[ii)] Models based on a physics-driven geometry-based prior $p(\bm{h}|\bm{\theta},\bm{\alpha})$ where priors of geometry-related channel parameters $p(\bm{\theta})$ and amplitudes $p(\bm{\alpha})$ are inferred and tracked online to predict future \gls{csi}~\cite{Palleit11Prediction,Paiva25Prediction}.
    \end{enumerate}%
%\end{enumerate}%
%\columnbreak
\newpage
%\begin{enumerate}[leftmargin=15pt]
    \item[B.] Data-driven approaches: 
    %\vspace{-5pt}%
    %\noindent B.~Data-driven approaches%
    \begin{enumerate}[leftmargin=2pt]
        \item[i)] Models based on a geometry-agnostic channel prior $p(\bm{h})$ learned offline from training data where geometry-related parameters enter only implicitly~\cite{Kim21KFvsMachineLearning,Thoota23dataDrivenBF,Rizzello24ChannelPrediction}.
        \item[ii)] Models based on a data-driven geometry-based prior $p(\bm{h}|\bm{\theta})$
        with parameter prior $p(\bm{\theta})$ learned offline from the training data~\cite{Neumann18MMSElearning,Fesl24DBchannelPrior}.
    \end{enumerate}%
\end{enumerate}%

\noindent\null%
{%
\setlength{\parindent}{0pt}%
This work belongs to category A.ii) where we devise a direct multipath-based \gls{slam} method~\cite{LiaLeiMey:Asilomar2023,LiaLeiMey:TSP2025} that works directly on the radio signal (the observed \gls{csi}) to infer the multipath channel parameters $\bm{\theta}$ and amplitudes $\bm{\alpha}$.
State-of-the-art multipath-based \gls{slam} methods assume an unknown model-order (i.e., number of sources $K$) and can be implemented in either a two-step approach, using a parametric channel estimation~\cite{HansenSAM2014,HanFleuRao:TSP2018,GreLeiWitFle:TWC2024,MoePerWitLei:TSP2024,MoeWesVenLei:Arxiv2025} algorithm followed by a multipath-based \gls{slam} algorithm ~\cite{GentnerTWC2016,LeitMeyHlaWitTufWin:TWC2019,KimGraGaoBatKimWym:TWC2020,KimGranSveKimWym:TVT2022, Leitinger23mvaSLAM,LeiWieVenWit:Asilomar2024,LiCaiLeiTuf:ICC2024}, or in a direct approach~\cite{LiaLeiMey:Asilomar2023,LiaLeiMey:TSP2025,FasDeuKesWilColWitLeiSecWym:STSP2025}. %
}
Leveraging the parameters inferred on the \gls{ul},
our method demonstrates efficient \gls{csi} predictions on the \gls{dl} under low \gls{snr} and user mobility, as well as robust positioning and \gls{csi} fusion even under \gls{los} obstructions.

\ifthenelse{\equal{\shortPaperVersion}{true}}%
{%      Omit the notation in the short paper
}{%
{\slshape Notation.} Lowercase bold letters $\bm{x}$ will be used to denote vectors, while uppercase bold letters $\bm{X}$ denote matrices.
Further, $\left[ \bm{X}\right]_{i,j}$ denotes the element of row $i$ and column $j$ in matrix $\bm{X}$. 
We use $\bm{x}^\trp$ and $\bm{x}^\herm$ to denote the transpose and Hermitian transpose of $\bm{x}$, respectively.
The Euclidean norm ($2$-norm) of a vector $\bm{x}$ is denoted as $\lVert \bm{x} \rVert$.
The magnitude of a complex number $z$ is denoted by $|z|$, and $z^*$ is the complex conjugate of $z$. 
The $N\!\times\!N$ identity matrix is denoted by $\eye{N}$, while $\onematrix{i}{j}$ denotes a $(i\!\times\!j)$-matrix of all ones. 
With $\bm{X}$ being an $(M\!\times\!N)$ matrix, $\bm{X}^\dag$ denotes its $(N\!\times\!M)$ Moore-Penrose pseudoinverse, and $|\bm{X}|$ its determinant.
We define the multivariate Dirac delta function as $\delta(\bm{x}):=\prod_{\scriptscriptstyle i=1}^{\scriptscriptstyle \dim(\bm{x})}\delta([\bm{x}]_i)$. 
}

\def\datapath{./figures}
\begin{figure}
\setlength{\figurewidth}{0.99\columnwidth}
    \centering
    \input{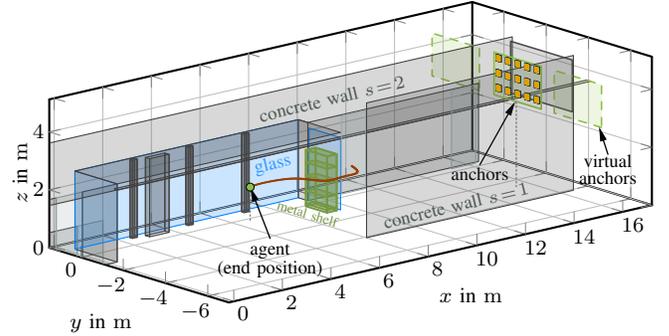}
    \caption{The measurement scenario: A mobile agent is moving on a trajectory around a shelf from \acrshort{los} to \acrshort{olos} conditions. %
    }
    \label{fig:TrajectoryScenario}
    \vspace{-0.5cm}
\end{figure}

\section{\MakeUppercase{Scenario, Method, and Hypotheses}}\label{sec:scenario}

%\paragraph{Scenario} 
{\slshape Scenario.} 
We perform i) \gls{ul} positioning and mapping and ii) \gls{dl} beamforming, where we evaluate our methods in a real-world hallway scenario schematically depicted in Fig.\,\ref{fig:TrajectoryScenario}. 
We conduct synthetic aperture measurements with a mechanical positioner where we subsequently measure the channels between a mobile single-antenna agent and $\Nrw\!=\!15$ static anchors, each equipped with an $(8\!\times\!8)$-\gls{ura}, that jointly form a \textit{physically large} aperture, i.e., an aperture that is large w.r.t. the propagation distances of interest~\cite{Wilding23Propagation}.
The anchors are assumed to be frequency synchronized through a shared reference clock and phase calibrated.
\ifthenelse{\equal{\shortPaperVersion}{true}}%
{%
    A Rohde\,\&\,Schwarz ZVA24 \gls{vna} is used to measure the transmission coefficients, i.e., \acrshortpl{s-parameter}, between a receiving anchor antenna $m$ and the transmitting agent antenna~\cite{Deutschmann23ICC}. 
}{%
    A Rohde\,\&\,Schwarz ZVA24 \gls{vna} is used in a two-port configuration to subsequently measure the transmission coefficients, i.e., \acrshortpl{s-parameter}, $S_{\scriptscriptstyle21,m}(f)$ % \triangleq \left[ \bm{h}(f)\right]_{\scriptstyle m}$ 
    between a receiving anchor antenna $m$ connected to Port\,$2$ and the transmitting agent antenna connected to Port\,$1$. 
    Linear systematic measurement errors introduced by cables and connectors are removed by a \gls{tosm} calibration which effectively shifts the measurement reference planes to the antenna ports. 
    We further deembed the electrical length of the used \gls{xets} antennas~\cite{Costa2009XETS} at the anchors and the used cent antenna~\cite[App.\,B.3.]{Krall08CentAntenna} at the agent.
}%
We emulate to operate with a bandwidth of $B\!=\!\SI{500}{\mega\hertz}$ centered around $\fc\!=\!\SI{6.175}{\giga\hertz}$
which matches with the \gls{nr} band 
\ifthenelse{\equal{\shortPaperVersion}{true}}%
{%
    n102, 
}{%
    n102~\cite{ETSI_TS_138_101}, 
}%
used for Wi-Fi\,6E defined in IEEE\,Std.\,802.11ax%\texttrademark
\,\cite{IEEE_802-11ax}.

%The mobile agent moves along the trajectory from strong \gls{los} conditions into total \gls{olos} conditions.
%Emulating a real-world scenario, 
%\paragraph{Problem Statement} {\color{IEEEred}Rename this paragraph or remove title}
%\paragraph{Method} %Our synthetic aperture measurements emulate a real-world problem where 
{\slshape Method.} 
The mobile agent transmits \gls{ul} pilots while moving around a shelf filled with highly absorbing\footnote{Pyramidal absorbers are used to cause a strong obstruction.} material from strong \gls{los} conditions into totally \gls{olos} conditions. 
On the \gls{ul}, the $\Nrw$ anchors acquire noisy channel observations---\textit{estimated} \gls{csi}---which are used in our geometry-based \textit{parametric} channel estimator to jointly infer the agent position and environment map.
On the \gls{dl}, efficient beamforming requires high-quality \gls{csi}.
Conjugate beamforming~\cite{mMIMObook} given estimated \gls{csi}, i.e., \textit{reciprocity-based} beamforming, suffers both from i) \textit{noisy} \gls{csi} estimates and from ii) \gls{csi} \textit{aging} due to agent mobility. 

{\slshape Hypotheses.}%
\begin{enumerate}[leftmargin=16.5pt]
    \item[\circleblue{1}] \textit{Robustness}: Robust and accurate positioning of the agent even in \gls{olos} conditions is enabled through \gls{nlos} components, i.e., rays that are reflected off walls, leveraging the previously learned environment map.
    %As the agent enters \gls{olos} conditions, the environment map, i.e., the learned positions and orientations of walls, shall be used to robustly position the agent using \gls{nlos} components, i.e., rays that are reflected off walls.
    \item[\circleblue{2}] \textit{Efficiency}: After being used for inference in the inverse problem of estimating geometric parameters from noisy estimated \gls{csi}, we leverage our geometry-based channel model to solve the forward problem of \textit{predicting} \gls{csi} given the inferred agent position and environment map. 
    Our geometry-based beamformer given predicted \gls{csi} can outperform a reciprocity-based beamformer given noisy \gls{csi} because it incorporates physics-based channel knowledge and rejects physically implausible \gls{csi}.
    \item[\circleblue{3}] \textit{Mobility Support}: We leverage a motion model to predict future agent positions and further predict \gls{csi} to future %(even non-integer) 
    time steps, which gives our geometry-based beamformer the ability to perform efficiently even under user mobility where a conventional reciprocity-based beamformer given outdated \gls{csi} suffers from losses due to \gls{csi} aging.
\end{enumerate}

\section{\MakeUppercase{Signal Model}}

\subsection{\MakeUppercase{Channel Model}}
We assume that at each time step $n \in \{1 \, \dots \, N\}$, each anchor $j \in \{1 \, \dots \, \Nrw\}$ acquires a noisy frequency-domain channel estimate
\begin{align}\label{eq:gen-signal-model}
        \hestimate{\nantenna}{n}{j} =
        \sum\limits_{\MVApair{s}{s'}\in\setSDBt} 
        \hcomponent{\nantenna}{n}{j}{s}{s'} +
        \bm{w}_{\scriptscriptstyle \nantenna,n,j}^{\text{\tiny AWGN}} \quad \in \complexset{\Nfrequency}{1}
\end{align}
per antenna $m \in \big\{1 \, \dots \, M\big\}$ in complex baseband with elements $[\widetilde{\bm{h}}_{\scriptscriptstyle \nantenna,n,j}]_{\nfrequency}$ and $\nfrequency \in \big\{-\!\frac{\Nfrequency-1}{2}  \, \dots \, \frac{\Nfrequency-1}{2} \big\}$ being discrete-frequency samples equally-spaced at $\Delta f \!=\! \frac{B}{\Nfrequency}$ and the number of frequency bins $\Nfrequency$ either being an odd integer, or $\nfrequency$ being centered between two integers.
Eq.\,\eqref{eq:gen-signal-model} consists of two terms: 
The first term represents a sum of a \gls{los} channel vector $\hcomponent{\nantenna}{n}{j}{0}{0}$ and up to $\Nwalls^2$ \gls{smc} channel vectors $\{\hcomponent{\nantenna}{n}{j}{s}{s'} \, |\, \MVApair{s}{s'}\in \setSDB\}$ modeling single-bounce reflections $\MVApair{s}{s}\in \setSB := \{\MVApair{s}{s} \in {\setWalls \times \setWalls}\}$ or double-bounce reflections $\MVApair{s}{s'}\in \setDB := \{\MVApair{s}{s'} \in {\setWalls \times \setWalls}|\,s\neq s'\}$ at large, planar surfaces $s \in \setWalls := \{1 \dots \Nwalls\}$, where $\setSDB \!:=\! \setSB \cup \setDB$ and $\setSDBt\!:=\!\MVApair{0}{0} \cup \setSDB$. 
The second term represents an \gls{awgn} vector of $\Nfrequency$ i.i.d. circular Gaussian noise samples $[\bm{w}_{\scriptscriptstyle \nantenna,n,j}^{\text{\tiny AWGN}}]_{\nfrequency} \sim \mathcal{CN}(0,\sigma_{\scriptscriptstyle j}^2)$.
Our inference channel model neglects \gls{dm} which typically represents stochastic scattering at small objects, or surfaces that are rough (w.r.t. the wavelength $\lambda$)~\cite{KulmerPIMRC2018,WieVenWilLei:JAIF2023,WieVenWilWitLei:Fusion2024}.

The $\Nantennas$ noisy channel estimates are stacked into a matrix $\widetilde{\bm{H}}_{\scriptscriptstyle n,j}\!=\!\big[ \widetilde{\bm{h}}_{\scriptscriptstyle 1,n,j} \, \dots \, \widetilde{\bm{h}}_{\scriptscriptstyle M,n,j} \big]\!\in\!\complexset{\Nfrequency}{\Nantennas}$ and vectorized to obtain the channel observation vector $\observation{n}{j}\!=\!\vect\!\big(\widetilde{\bm{H}}_{\scriptscriptstyle n,j} \big) \in \complexset{\Nfrequency\,\Nantennas}{1}$, which can be expressed in matrix-vector notation as 
\begin{align}\label{eq:observation}
    \observation{n}{j} =
    \dictionary{j}(\state{n}) 
    \alphavec{n,j} 
    + \noise{n}{j} \,,
\end{align}
assuming temporally and spatially uncorrelated noise $\noise{n}{j}\sim \mathcal{CN}(\bm{0},\sigma_{\scriptscriptstyle j}^2 \eye{\Nfrequency\Nantennas})$, and where $\dictionary{j}(\state{n})\!=\![\dictEntry{0}{0}\!\dots\,\dictEntry{s'}{s'} ]\!\in\!\complexset{\Nfrequency \Nantennas}{\Ncomponents}$, parameterized\footnote{The dependence of $\dictionary{j}$ on $\state{n}$ is frequently omitted for notational brevity.} by a (random) state vector $\state{n}$ capturing geometry-related parameters, is a dictionary matrix with $\Ncomponents \!:=\! |\setSDBt|\!=\!S^2\!+\!1$ column-vectors $\dictEntry{s}{s'}\!(\state{n})\!=\!\vect\!\big({\bm{H}}_{\scriptscriptstyle n,j}^{\MVApair{s}{s'}} \big)$. 
They are found as the vectorizations of the spatiotemporal array manifold ${\bm{H}}_{\scriptscriptstyle n,j}^{\MVApair{s}{s'}}\!\in\!\complexset{\Nfrequency}{\Nantennas}$ with unit-modulus elements
\begin{align}\label{eq:spatiotemporal-array-manifold}
    \left[{\bm{H}}_{\scriptscriptstyle n,j}^{\MVApair{s}{s'}} \right]_{\nfrequency,\nantenna} =%\big( d_{\scriptscriptstyle m,n}^{\MVApair{s}{s'}} \big) = 
    \exp \left(
        -\mathrm{j} \frac{2\pi}{\lightspeed} (\fc + \Delta f \, \nfrequency) \, d_{\scriptscriptstyle m,n,j}^{\MVApair{s}{s'}}
    \right)
\end{align}
parameterized on the path lengths $d_{\scriptscriptstyle m,n,j}^{\MVApair{s}{s'}}(\state{n})$ that are a function of the state vector.
The amplitudes $[\alphavec{n,j}]_\ncomponent$ for one component $\ncomponent\!\in\!\{1 \dots \Ncomponents\}$ %$\MVApair{s}{s'}$ 
in~\eqref{eq:observation} are assumed to be approximately constant over all antennas $m$ of one anchor $j$ and capture the lumped effects of direction-dependent antenna gains, distance-dependent path losses, losses due to impedance and polarization mismatch, as well as losses incurring due to reflections at specular surfaces~\cite{Deutschmann23ICC,Deutschmann2025WCM}.
The important link between the array manifold and geometry is captured in the path lengths $d_{\scriptscriptstyle m,n,j}^{\MVApair{s}{s'}}$ described by our geometric model, which we define next and which leads to our choice of $\state{n}$.

\subsection{\MakeUppercase{LoS and Specular Multipath Model}}

To describe the environment geometry, we use the \gls{mva} model from~\cite{Leitinger23mvaSLAM} defining a specular surface $s$ solely through the \gls{mva} position $\pmva{s}$, which is computed by mirroring the origin $\origin$ of the global Cartesian coordinate system across surface $s$.
It elegantly represents both the position of the surface through a wall-point 
\begin{align}\label{eq:pwk}
    \pwk = \frac{\pmva{s}}{2} \qquad \in \mathbb{R}^{3 \times 1}
\end{align}
and the surface orientation through a normal vector
\begin{align}\label{eq:nw}
    \nw = \frac{\pmva{s}}{\lVert \pmva{s} \rVert} \qquad \in  \mathbb{R}^{3 \times 1}
\end{align}
with only a single variable $\pmva{s} \in \mathbb{R}^{3 \times 1}$.

\subsubsection{Physical Anchors}\noindent
We define \glspl{pa} $\nsubarray$ centered around the phase center position $\posAnchor{j}{0}{0}\!\in\!\realset{3}{1}$ with antennas $\nantenna$ %$m \in \{1 \, \dots \, M \}$ 
located at positions $\posantenna{m}\!\in\!\realset{3}{1}$ \textit{relative} to $\posAnchor{j}{0}{0}$.
The complete ``template'' anchor array layout relative to $\posAnchor{j}{0}{0}$ is captured in $\Ptemplate = [ \posantenna{1} \, \dots \, \posantenna{M} ] \in \realset{3}{M}$ and is identical for all anchors. 
The antenna positions of an actual \textit{physical} anchor are computed by shifting a ``\textit{template}'' anchor out of the origin $\origin$ to its center position $\posAnchor{j}{0}{0}$ in global coordinates through
\begin{align}\label{eq:PA-layout}
    \Pjss{j}{0}{0} =  \posAnchor{j}{0}{0} \, \onematrix{1}{M}\, + \rotM \Ptemplate \quad \in \realset{3}{M} ,
\end{align}
where $\onematrix{i}{j}$ denotes an $(i\times j)$-matrix of all ones. 
The matrix $\rotM \in \realset{3}{3}$ is a rotation matrix that defines the orientation of \gls{pa} $j$ in global coordinates.

\ifthenelse{\equal{\shortPaperVersion}{true}}%
{
    % skip in short paper
}{
    \begin{figure}
    %\begin{minipage}[t]{0.45\linewidth}
        \vskip 0pt	% NECESSARY TO MAKE THE TOP "[t]" PARAMETER WORK!
        \setlength{\plotWidth}{1.4\linewidth}
        \hspace{-1cm}\vspace{-0.75cm}\input{figures/P3343-SMC-model}~~
        \vspace{-1cm}\captionof{figure}{
        The array layout of the \acrshort{pa} is computed using \eqref{eq:PA-layout} while the layout of a single-bounce \acrshort{va} is computed using~\eqref{eq:VA-layout-SB}. 
        An \acrshort{mva} position $\pmva{s}$ is used as an elegant description of a specular surface $s$ which defines its normal vector $\nw$ through~\eqref{eq:nw} and an arbitrary point $\pwk$ on the surface through~\eqref{eq:pwk}.}
        \label{fig:P3343-SMC-model}
    %\end{minipage}
    \end{figure}
}%

\subsubsection{Single-Bounce Virtual Anchors}\noindent
%\Glspl{smc} 
Reflections of the agent's \gls{ul} pilots at large specular surfaces are modeled as if they were virtually impinging at \glspl{va} that are images of \glspl{pa} mirrored across surfaces $s$%
\ifthenelse{\equal{\shortPaperVersion}{true}}%
{. % skip in short paper
}{% reference in long paper
    (see Fig.\,\ref{fig:P3343-SMC-model}). 
}%.
We model \glspl{va} using \glspl{mva} as described in~\cite{Leitinger23mvaSLAM}:
The transformation from a \gls{pa} to a \gls{va} phase center position is computed using the function $\hva : \mathbb{R}^3 \times \mathbb{R}^3 \to \mathbb{R}^3$
defined as
\begin{align}\label{eq:posAnchor-SB}
    \posAnchor{j}{s}{s} &= \hva(\posAnchor{j}{0}{0},\pmva{s})  \\ \nonumber
    &= \posAnchor{j}{0}{0} -\left( 
        \frac{2 \, \posAnchor{j}{0}{0}^\trp \pmva{s}}{\lVert \pmva{s} \rVert^2} - 1
    \right) \, \pmva{s} \, .
\end{align}
The Householder matrix
\begin{align}\label{eq:house}
    \house{s} = \eye{3} - 2 \, \frac{{\pmva{s}} {\pmva{s}}^\trp}{\lVert \pmva{s} \rVert^2}
\end{align}
represents the transformation from the \gls{pa} orientation to the \gls{va} orientation when mirrored across specular surface $s$.
The complete array layout of a single-bounce \gls{va} is captured in
\begin{align}\label{eq:VA-layout-SB}
    \Pjss{j}{s}{s} =  \posAnchor{j}{s}{s} \, \onematrix{1}{M}\, + \house{s} \rotM \Ptemplate  \quad \in \realset{3}{M}\, .
\end{align}

\subsubsection{Double-Bounce Virtual Anchors} \noindent
As described in~\cite{Leitinger23mvaSLAM}, a double-bounce \gls{va} phase center position %a \gls{pa} position 
is computed by applying~\eqref{eq:posAnchor-SB} twice, i.e., $\posAnchor{j}{s}{s'} = \hva(\hva(\posAnchor{j}{0}{0}, \pmva{s}), \pmva{s'})$. 
The orientation transformation of the double bounce path $\MVApair{s}{s'}$ is taken into account by applying Householder matrices $\house{s}$ and $\house{s'}$ from both surfaces using~\eqref{eq:house}, which leads to the representation of the complete array layout of a double-bounce \gls{va} in global coordinates as
\begin{align}\label{eq:VA-layout-DB}
    \Pjss{j}{s}{s'} =  \posAnchor{j}{s}{s'} \, \onematrix{1}{M}\, + \house{s'} \house{s} \rotM \Ptemplate \quad \in \realset{3}{M}\, .
\end{align}
Linking our geometric model to the array manifold in~\eqref{eq:spatiotemporal-array-manifold}, the path lengths 
\begin{align}
    d_{\scriptscriptstyle m,n,j}^{\MVApair{s}{s'}} = \lVert \bm{p}_{\scriptscriptstyle m,j}^{\MVApair{s}{s'}} - \pos{n}\rVert
\end{align}
are the scalar distances from the agent at $\pos{n}\in\realset{3}{1}$ to the receiving antenna positions $\bm{p}_{\scriptscriptstyle m,j}^{\MVApair{s}{s'}}$ captured\footnote{Position $\bm{p}_{\scriptscriptstyle m,j}^{\MVApair{s}{s'}}$ is the $m$\textsuperscript{th} column vector in the respective layout.} in the layout for \glspl{pa} in~\eqref{eq:PA-layout}, for single-bounce \glspl{va} in~\eqref{eq:VA-layout-SB}, and for double-bounce \glspl{va} in~\eqref{eq:VA-layout-DB}.

\subsection{Problem Formulation}
In the \textit{inverse} problem, we aim to infer the geometry-based channel parameters, i.e., the state $\state{n}$, that led to the data, i.e., the %\gls{dmimo} 
anchor infrastructure observations $\observationMatrix{n}\!:=\!\big[\observation{1}{n} \, \dots \, \observation{\Nrw}{n}\big]\!\in\!\complexset{\Nfrequency \Nantennas}{\Nrw}$.
We choose a joint state vector 
\begin{align}\label{eq:state-vector}
    \state{n} = \big[\agentState{n}^\trp, ~{\PMVAposStacked{n}}^\trp\big]^\trp \quad \in \stateSpace \, ,
\end{align}
with $\stateSpace\!=\!\mathbb{R}^{5 + 3\Nwalls}$ denoting the state space, which comprises the agent state $\agentState{n}$ and the stacked \gls{mva} state $\PMVAposStacked{n}$.
The agent state $\agentState{n}\!=\![\pos{n}^\trp, \velocity{n}^\trp]^\trp\!\in\!\realset{5}{1}$ contains the three-dimensional agent position $\pos{n}\!\in\!\realset{3}{1}$\!, but only a horizontal agent velocity $\velocity{n}\!\in\!\realset{2}{1}$ as the agent has no vertical motion.
Capturing the map information, the stacked \gls{mva} state $\PMVAposStacked{n} = \big[{\pmva{1}}^\trp \,  \dots \, {\pmva{S}}^\trp\big]^\trp\!\in\!\realset{3\Nwalls}{1}$ solely comprises the \gls{mva} positions in this work, although it could be extended to track the existence %\footnote{This allows to model obstructions and limited surface extents.} 
of a surface~\cite{Leitinger23mvaSLAM}, or parameters describing its reflective properties.

In the \textit{forward} problem, we aim to predict the data, i.e., a %geometry-based 
channel vector $\hpredict{n,j}$, parameterized by our estimated parameters $\stateEstimate{n}$.
Our methods to solve the inverse and forward problems are found in Sec.\,\ref{sec:tracking} and Sec.\,\ref{sec:est-pred-fusion}, respectively.

\section{\MakeUppercase{Spatial-Delay Positioning and Mapping}}\label{sec:tracking}%

To fuse the information on the state %$\state{n}$ 
acquired on subsequent observations, we employ 
the Bayesian filtering equation~\cite{Gustafsson10PFtheory}
\begin{align}\label{eq:posterior}
    p(\state{n}|\observationMatrix{1:n}) = \frac{p(\observationMatrix{n}|\state{n}) p(\state{n}| \observationMatrix{1:n-1})}{p(\observationMatrix{n}| \observationMatrix{1:n-1})}
\end{align}
that describes the posterior \gls{pdf} $p(\state{n}|\observationMatrix{1:n})$ of the state vector $\state{n}$ at time step $n$ given past observations (i.e.,\,measurements) $\observationMatrix{1:n-1}\!:=\!\{\observationMatrix{1}, \dots , \observationMatrix{n-1}\}$ and the current observation $\observationMatrix{n}$.
Assuming a first-order Markov chain, i.e., $p(\state{n}|\state{n-1}\,\dots \, \state{0})\!=\!p(\state{n}|\state{n-1})$, the prediction \gls{pdf} $p(\state{n}|\observationMatrix{1:n-1})$ is computed from the state-transition \gls{pdf} and the previous posterior \gls{pdf} through the Chapman-Kolmogorov equation
\begin{align}
    p(\state{n}|\observationMatrix{1:n-1})\!=\!\int_{\stateSpace} p(\state{n}|\state{n-1},\observationMatrix{1:n-1}) p(\state{n-1}| \observationMatrix{1:n-1}) \diff \state{n-1} \nonumber
\end{align}
where we assume that the state-transition \gls{pdf} $p(\state{n}|\state{n-1},\observationMatrix{1:n-1})\!=\!p(\state{n}|\state{n-1})\!:=\!\mathcal{N}(\transitionmatrix \state{n-1},\processNoiseCov)$ is defined through the state-space model 
\begin{align}\label{eq:state-space-model}
    \state{n} = \transitionmatrix \state{n-1} + \processNoise{n}
\end{align}
where $\processNoise{n}\!\in\!\stateSpace$ denotes zero-mean Gaussian process noise with a diagonal covariance matrix $\processNoiseCov\!:=\! \diag \big(\sigma^2_{\scriptscriptstyle \mathrm{p}}, \sigma^2_{\scriptscriptstyle \mathrm{p}}, \sigma^2_{\scriptscriptstyle \mathrm{p}}, \sigma^2_{\scriptscriptstyle \mathrm{v}}, \sigma^2_{\scriptscriptstyle \mathrm{v}}, \sigma^2_{\text{\tiny mva}} \, \dots \,  \sigma^2_{\text{\tiny mva}} \big)$ and $\transitionmatrix$ is a state-transition matrix defined as
\begin{align}
    \left[\transitionmatrix\right]_{\scriptscriptstyle k,\ell} = 
    \begin{cases}
        1 & k=\ell \\
        \Delta t & \big((k,\ell) = (1,4)\big) \lor \big((k,\ell) = (2,5)\big) \\
        0 & \text{else},
    \end{cases}
\end{align}
with $\Delta t$ denoting the time interval between two subsequent observations $\observationMatrix{n-1}$ and $\observationMatrix{n}$.
In Sec.\,\ref{sec:likelihood-models} we define two models for the likelihood $p(\observationMatrix{n}|\state{n})$ of the observation conditional on the state.

We approximately implement~\eqref{eq:posterior} through a~\gls{pf} with a \textit{random measure} $\{\particle{n}{i},\weight{n|n}{i}\}_{i=1}^{\Nparticles}$ of $\Nparticles$ particles $\particle{n}{i}$ with weights $\weight{n|n}{i}$ s.t. $\sum_{i=1}^{\Nparticles}\weight{n|n}{i}\!\triangleq\!1$, implicitly accounting for the normalization constant $p(\observationMatrix{n}| \observationMatrix{1:n-1})$. 
The \gls{pf} approximates the posterior~\gls{pdf} as $\hat{p}(\state{n}|\observationMatrix{1:n})\!=\!\sum\nolimits_{i=1}^{\Nparticles} \weight{n|n}{i} \delta \big( \state{n}\!-\particle{n}{i} \big)$ from which we estimate the state $\state{n}$ by approximating the \gls{mmse} estimate 
$\state{n}^{\text{\tiny MMSE}}=\mathbb{E}\left(\state{n}|\observationMatrix{1:n}\right)$ as
\begin{align}\label{eq:state-estimate}
    \stateEstimate{n} = \int_{\stateSpace}\!\state{n} \,\hat{p}(\state{n}|\observationMatrix{1:n}) \,  \diff \state{n}= \sum\limits_{i=1}^{\Nparticles}  \particle{n}{i} \weight{n|n}{i} \, ,
\end{align}
and approximate the state covariance matrix as
\begin{align}\label{eq:empiricalCovariance}
    \covarianceEstimate{n} =  \sum\nolimits_{i=1}^{\Nparticles}
    \left(\particle{n}{i} - \stateEstimate{n}\right)\left(\particle{n}{i} - \stateEstimate{n}\right)^{\!\trp} \weight{n|n}{i} \, .
\end{align}
We use a regularized~\gls{pf}, implemented through~\cite[Alg.\,1]{Deutschmann24SPAWC} by replacing the likelihood function in line~6 with the stochastic likelihood function implemented by Alg.\,\ref{alg:sto-lhf} below.

\section{\MakeUppercase{Likelihood Model}}\label{sec:likelihood-models}

\subsection{\MakeUppercase{Deterministic Concentrated Likelihood}}
The statistical model of the observations $\observation{n}{j}$ in~\eqref{eq:observation} underlying the deterministic concentrated likelihood~\cite{KrimViberg96ASP} assumes that the %source 
signal amplitudes $\alphavec{n,j}$ are \textit{deterministic} unknowns.
Per~\eqref{eq:observation}, the per-anchor observation is distributed as $\observation{n}{j}|\state{n}\!\sim\!\mathcal{CN}(\dictionary{j}(\state{n}) \alphavec{n,j}, \sigma_{\scriptscriptstyle j}^2 \eye{\Nfrequency\Nantennas})$.
Assuming independent observations among all anchors $j$ leaves us with the deterministic likelihood function %which is the \gls{pdf}
\begin{equation}
    \begin{split}\label{eq:lhf-det}
    &p\big(\observationMatrix{n};\state{n},
    \{\alphavec{n,j}\}_{\scriptscriptstyle j=1}^{\scriptscriptstyle J}, 
    \{\sigma_{\scriptscriptstyle j}^2\}_{\scriptscriptstyle j=1}^{\scriptscriptstyle J}\big) 
    \\ 
    & = \prod_{j=1}^{J}
    \frac{
        \exp \left( - \frac{1}{\sigma_{\scriptscriptstyle j}^2} \left\| \observation{n}{j} - \dictionary{j}(\state{n}) \alphavec{n,j} \right\|^2\right)
    }{
        \left(\pi \sigma_{\scriptscriptstyle j}^2\right)^{\Nfrequency\Nantennas}
    } %\nonumber
    \end{split}
\end{equation}
for the infrastructure observation $\observationMatrix{n}$ parameterized by the state $\state{n}$, the amplitudes $\{\alphavec{n,j}\}_{\scriptscriptstyle j=1}^{\scriptscriptstyle J}$, and noise variances $\{\sigma_{\scriptscriptstyle j}^2\}_{\scriptscriptstyle j=1}^{\scriptscriptstyle J}$.
Since the latter two contribute negligible information about the state $\state{n}$, we treat them as nuisance parameters and we compute the \textit{profile} likelihood, i.e., we \textit{concentrate} w.r.t. the nuisance parameters by computing \gls{ml} estimates conditional on the state $\state{n}$.
\Gls{ml} estimates of amplitudes are found as 
\begin{align}\label{eq:det-alpha-hat}
    \alphavecHat{n,j} | \state{n} = \dictionary{j}^\dag%(\state{n})
    \, \observation{n}{j} \, ,
\end{align}
with $\dictionary{j}^\dag(\state{n})$ denoting the pseudoinverse of the dictionary, while \gls{ml} estimates of noise variances are computed as~\cite[eq.\,(48)]{KrimViberg96ASP}
\begin{align}\label{eq:det-sigma2-hat}
    \sigmaHat{j} | \state{n} = 
    \frac{1}{\Nfrequency \Nantennas} \tr \left(
    \PiPerp{j} \Rhate{n}{j}
    \right)
\end{align}
with $\PiPerp{j}(\state{n})\!=\!\eye{\Nfrequency\Nantennas}\!-\!\dictionary{j}\dictionary{j}^\dag$ being the projector onto the noise subspace, i.e., the orthogonal complement of the subspace spanned by $\dictionary{j}(\state{n})$, and the sample covariance matrix $\Rhate{n}{j}\!:=\!\observation{n}{j} \, {\observation{n}{j}}^\herm %\in \complexset{\Nfrequency\Nantennas}{\Nfrequency\Nantennas}
$ is an \textit{empirical} rank-$1$ estimate of the spatiotemporal signal covariance matrix $\signalCov{n,j}=\mathbb{E}\big(\observation{n}{j} {\observation{n}{j}}^\herm\big)$.
Despite notationally brief,~\eqref{eq:det-sigma2-hat} %and~\eqref{eq:lhf-det-profile} are 
is computationally expensive and can be implemented efficiently using~\eqref{eq:efficient-trace-term} %and~\eqref{eq:efficient-likelihood} 
in Appendix~\ref{app:implementation}.
Reinsertion of $\alphavecHat{n,j} | \state{n}$ and $\sigmaHat{j} | \state{n}$ in~\eqref{eq:lhf-det} yields the deterministic profile likelihood function\footnote{Note that $\observation{n}{j}\!-\!\dictionary{j}\alphavecHat{n,j}\triangleq \PiPerp{j} \observation{n}{j}$ and~\eqref{eq:efficient-trace-term} omit computing $\PiPerp{j}$.}
\begin{align}\label{eq:lhf-det-profile}
    p(\observationMatrix{n}|\state{n}) 
    = %\\ \nonumber
    \prod_{j=1}^{J}
    \frac{
        \exp \left( - \frac{1}{\sigmaHat{j}} \left\| \PiPerp{j} \observation{n}{j} %- \dictionary{j}(\state{n}) \alphavecHat{n,j} 
        \right\|^2\right)
    }{
        \left(\pi \sigmaHat{j}\right)^{\Nfrequency\Nantennas}
    }\,.
\end{align}

\subsection{\MakeUppercase{Stochastic Concentrated Likelihood}}\label{sec:sto-lhf}

The statistical model of the observations $\observation{n}{j}$ in~\eqref{eq:observation} underlying the stochastic concentrated likelihood~\cite{Stoica1995} assumes that the signal amplitudes are \textit{stochastic} unknowns with prior \gls{pdf} $p(\alphavec{n,j}|\mualpha{n,j},\sourceCov{n,j})\!=\!\mathcal{CN}(\mualpha{n,j},\sourceCov{n,j})$ parameterized by an unknown source covariance matrix $\sourceCov{n,j}\!\in\!\complexset{\Ncomponents}{\Ncomponents}$. 
Contrary to the established definition in the literature, we define amplitudes to have an unknown mean $\mualpha{n,j}\!\neq\!\bm{0}$, yet in Appendix~\ref{app:unbiasedness} we show that the established \gls{ml} estimators are still unbiased in concentrating the likelihood in~\eqref{eq:lhf-sto} w.r.t. the nuisance parameters.
It follows per~\eqref{eq:observation} that the per-anchor observation is distributed as $\observation{n}{j}|\state{n}\!\sim\!\mathcal{CN}(\dictionary{j}(\state{n}) \mualpha{n,j}, \signalCov{n,j})$ with the spatiotemporal signal covariance matrix $\signalCov{n,j}\!=\!\dictionary{j} \sourceCov{n,j} \dictionary{j}^\herm + \sigma_{\scriptscriptstyle j}^2 \eye{\Nfrequency\Nantennas}$.
Assuming independent observations among all anchors $j$ leaves us with the stochastic likelihood function 
\begin{align}
    %\begin{split}
    \label{eq:lhf-sto}
    &p\big(\observationMatrix{n}|\state{n},
    \{\mualpha{n,j}\}_{\scriptscriptstyle j=1}^{\scriptscriptstyle J}, 
    \{\signalCov{n,j}\}_{\scriptscriptstyle j=1}^{\scriptscriptstyle J}
    \big) 
    \\ 
    & =\!\prod_{j=1}^{J}%
    \!\frac{
        \exp \left(\!%
        - \big( \observation{n}{j}\!-\!\dictionary{j}(\state{n}) \mualpha{n,j} \big)^{\!\herm}
            \signalCov{n,j}^{-1}
        \big( \observation{n}{j}\!-\!\dictionary{j}(\state{n}) \mualpha{n,j} \big)
        \right)
    }{
        \pi^{\Nfrequency\Nantennas}
        \left| \signalCov{n,j} \right|
    } \,\nonumber
\end{align}
for the infrastructure observation $\observationMatrix{n}$.
Again, we compute the profile likelihood by concentrating w.r.t. the nuisance parameters, using the \gls{ml} estimators (cf.\,\cite{KrimViberg96ASP,Stoica1995})
\begin{align}
    \alphavecHat{n,j} | \state{n} &= \dictionary{j}^\dag%(\state{n}) 
    \, \observation{n}{j} \, , \label{eq:sto-alpha-hat} \\
    \sigmaHat{j} | \state{n} &= 
    \frac{1}{\Nfrequency \Nantennas - \Ncomponents} \tr \left(
    \PiPerp{j} \Rhate{n}{j}
    \right) \, , \label{eq:sto-sigma2-hat} \\
    \Phat{n}{j} | \state{n} &= 
    \dictionary{j}^\dag 
    \left(
        \Rhate{n}{j} - \sigmaHat{j} \eye{\Nfrequency\Nantennas}
    \right)
    {\dictionary{j}^\dag}^\herm \, ,\label{eq:sto-P-hat} 
\end{align}
with which we define a refined estimate
$\Rhat{n}{j} | \state{n}\!:=\!\dictionary{j} \Phat{n}{j} \dictionary{j}^\herm+\sigmaHat{j} \eye{\Nfrequency\Nantennas}$ that, reinserted in~\eqref{eq:lhf-sto}, yields the stochastic profile likelihood function
\begin{align}\label{eq:lhf-sto-profile}
    p\big(\observationMatrix{n}|\state{n}\big) 
    & \!=\!\prod_{j=1}^{J}%
    \!\frac{
        \exp \left(\!%
        - \left\|
            \Rhat{n}{j}^{-\frac{1}{2}}
            \big( \observation{n}{j}\!-\!\dictionary{j}(\state{n}) \alphavecHat{n,j} \big) 
        \right\|^2
        \right)
    }{
        \pi^{\Nfrequency\Nantennas}
        \big| \Rhat{n}{j} \big|
    } %\,.
\end{align}
assuming $\Rhat{n}{j} | \state{n}$ to be positive definite.
Despite notationally brief,~\eqref{eq:lhf-sto-profile} 
is computationally expensive and can be implemented efficiently using~\eqref{eq:detR-eff} and~\eqref{eq:invR-eff} %and~\eqref{eq:efficient-likelihood} 
from Appendix~\ref{app:implementation}.
\SetAlgoSkip{0pt}% reduce margins
\begin{algorithm}[t] 
	\LinesNumbered		% numbering
	\caption{Concentrated Stochastic Likelihood}\label{alg:sto-lhf}
    \SetAlgoLined\DontPrintSemicolon
    \SetKwFunction{computeVA}{computeVA}
    \SetKwFunction{generateResponse}{psi}
	\SetKwInOut{Input}{Input}\SetKwInOut{Output}{Output}
	\Input{Observation $\observationMatrix{n}$, particle $\particle{n}{i} %, \Nrw, \Ncomponents
        $}
        \Output{Unnormalized weight $\widetilde{w}_{\scriptscriptstyle n|n}^{\scriptscriptstyle (i)} \gets p(\observationMatrix{n}|\particle{n}{i})$}
        $\pos{n} \gets [\particle{n}{i}]_{\scriptscriptstyle 1:3}$, $\PMVAposStacked{n}  \gets [\particle{n}{i}]_{\scriptscriptstyle 6:5+3\Nwalls}$, and $\widetilde{w}_{\scriptscriptstyle n|n}^{\scriptscriptstyle (i)} \gets 0$\;
	  \For{$j \gets 1$ \KwTo $\Nrw$ \KwBy $1$}{ 
                $\dictionary{j} \gets \bm{0}$ and $\ncomponent \gets 1$\;
                $[\dictionary{j}]_{\scriptscriptstyle :,\ncomponent} \gets \texttt{\small psi}(\pos{n},\posAnchor{j}{0}{0}\!,\Ptemplate,\eye{3},\eye{3})$\;
	  	  \For{$s \gets 1$ \KwTo $S$ \KwBy $1$}{ 
                    $\big\{\posAnchor{j}{s}{s}\!,\house{s}\big\}\gets \texttt{\small computeVA}(\posAnchor{j}{0}{0},\pmva{s})$\;
                    \For{$s' \gets 1$ \KwTo $S$ \KwBy $1$}{ 
                        $\ncomponent \gets \ncomponent+1$\;
                        \uIf{$s = s'$}{
                            $[\dictionary{j}]_{\scriptscriptstyle :,\ncomponent}\!\gets\!\texttt{\small psi}(\pos{n},\posAnchor{j}{s}{s}\!,\Ptemplate,\house{s},\eye{3})$\; 
                          }
                          \Else{
                            $\big\{\posAnchor{j}{s}{s'}\!,\house{s'}\big\}\!\gets\!\texttt{\small computeVA}(\posAnchor{j}{s}{s},\pmva{s'})$\;
                            $[\dictionary{j}]_{\scriptscriptstyle :,\ncomponent}\!\gets\!\texttt{\small psi}(\pos{n},\posAnchor{j}{s}{s'}\!,\Ptemplate,\house{s},\house{s'})$\; 
                          }
    	        }
    	    }
            $\Rhate{n}{j} \gets \observation{n}{j} \, {\observation{n}{j}}^\herm$\; 
            $\alphavecHat{n,j} | \particle{n}{i}  \gets  \dictionary{j}^\dag%(\particle{n}{i}) 
    \, \observation{n}{j}$ \hfill \textcolor{gray}{//see \eqref{eq:sto-alpha-hat}\hspace{-4mm}}\\
            $\sigmaHat{j} | \particle{n}{i}  \gets  \frac{1}{\Nfrequency \Nantennas- \Ncomponents} \tr \big(\PiPerp{j} \Rhate{n}{j}\big)$ \hfill \textcolor{gray}{//\eqref{eq:sto-sigma2-hat} using \eqref{eq:efficient-trace-term}\hspace{-4mm}}\\
            $\Phat{n}{j} | \particle{n}{i}  \gets  \dictionary{j}^\dag \big(\Rhate{n}{j} - \sigmaHat{j} \eye{\Nfrequency\Nantennas}\big){\dictionary{j}^\dag}^\herm$ \hfill \textcolor{gray}{//see \eqref{eq:sto-P-hat-efficient}\hspace{-4mm}}\\
            $\widetilde{w}_{\scriptscriptstyle n|n}^{\scriptscriptstyle (i)} \gets \widetilde{w}_{\scriptscriptstyle n|n}^{\scriptscriptstyle (i)} \times p(\observation{n}{j}|\particle{n}{i})$\hfill \textcolor{gray}{//\eqref{eq:lhf-sto} using \eqref{eq:detR-eff},\eqref{eq:invR-eff}\hspace{-4mm}}\\
        }
        \SetKwProg{myproc}{Sub-routine}{}{}
        \myproc{${\dictEntry{s}{s'}}\gets$\generateResponse{$\pos{n},\posAnchor{j}{s}{s'}\!,\Ptemplate,\house{s},\house{s'}$}}{
            \nl 
            $\Pjss{j}{s}{s'} \gets \posAnchor{j}{s}{s'} \, \onematrix{1}{M}\, + \house{s'} \house{s} \rotM \Ptemplate$\hfill \textcolor{gray}{//see~\eqref{eq:VA-layout-DB}\hspace{-4mm}}\\
            $\bm{d}\gets {\texttt{\small vecnorm}}\big(\Pjss{j}{s}{s'}\big)^\trp$\hfill \textcolor{gray}{//$(\Nantennas\times1)$ distances\hspace{-4mm}}\\
            ${\bm{H}}_{\scriptscriptstyle n,j}^{\MVApair{s}{s'}} \gets
            \exp \left(
                -\mathrm{j} \frac{2\pi}{\lightspeed} (\fc~ \onematrix{\Nfrequency}{1} + \bm{f})\,\bm{d}^\trp
            \right)$\hfill \textcolor{gray}{//see\,\eqref{eq:spatiotemporal-array-manifold}\textsuperscript{$\ast$}\hspace{-4mm}}\\
            $\dictEntry{s}{s'}\gets\vect\!\big({\bm{H}}_{\scriptscriptstyle n,j}^{\MVApair{s}{s'}} \big)$ \;
        }
        \myproc{$\{\posAnchor{j}{s}{s}\!,\house{s}\}\gets$\computeVA{$\posAnchor{j}{0}{0},\pmva{s}$}}{
            \nl 
            $\posAnchor{j}{s}{s} \gets \posAnchor{j}{0}{0} - (2 \posAnchor{j}{0}{0}^\trp \pmva{s} /\lVert \pmva{s} \rVert^2 \!-\!1) \pmva{s}$\hfill \textcolor{gray}{//see~\eqref{eq:posAnchor-SB}\hspace{-4mm}}\\
            $\house{s}\gets \eye{3}-2\pmva{s}{\pmva{s}}^\trp/\lVert \pmva{s} \rVert^2$\hfill \textcolor{gray}{//see~\eqref{eq:house}\hspace{-4mm}}\\
        %\nl \KwRet $\{\posAnchor{j}{s}{s}\!,\house{s}\}$\;
        }
     \vspace{-0.01cm}     
\algorithmfootnote{%
\textsuperscript{$\ast$}\footnotesize\,To be understood as element-wise extension of the exponential function with $\bm{f}\in\realset{\Nfrequency}{1}$ denoting the baseband frequency vector.} \vspace{-0.0cm}
\end{algorithm}%
\section{\MakeUppercase{Channel Estimation, Prediction, and Fusion}}\label{sec:est-pred-fusion}
In the forward problem, we aim to predict data, i.e., a channel vector $\hpredict{n,j}$, given the estimated parameters, i.e., the state estimate $\stateEstimate{n}$. 
In this work, we evaluate channel prediction only at the %(narrowband) 
carrier frequency $\fc$, i.e., we henceforth use $\Nfrequency\!=\!1$ for prediction.\footnote{We keep the notation of the preceding sections, assuming that the quantities are respectively scaled in dimensions.}
Under the assumption of frequency synchrony, we perform~\gls{cjt} with all $\Nrw$ anchors, although we discuss \gls{csi} estimation, prediction, and fusion only for a single anchor $j$.
For notational brevity, we omit the conditioning on $\observationMatrix{1:n}$ for the remainder of this section.

{\slshape Estimated CSI.} 
Stacking the $\Nantennas$ noisy channel estimates of anchor $j$ into a vector $\hmeas{n,j}\!=\!\big[ \widetilde{{h}}_{\scriptscriptstyle 1,n,j} \, \dots \, \widetilde{{h}}_{\scriptscriptstyle M,n,j} \big]^\trp\!\in\!\complexset{\Nantennas}{1}$, estimated \gls{csi} is obtained which satisfies 
$\hmeas{n,j}\!=\!\argmax_{\htrue{n,j}} \big( p(\hmeas{n,j} | \htrue{n,j}, \sigma_{\scriptscriptstyle j}^2) \big)$.

{\slshape Predicted CSI.} 
Concentrating the channel prior \gls{pdf} $p\big( \htrue{n,j} | \state{n}, \alphavec{n,j} \big)$ with $\alphavecHat{n,j} | \state{n}$ from~\eqref{eq:sto-alpha-hat} at the state estimate $\stateEstimate{n}$, we compute predicted \gls{csi} as
\begin{align}\label{eq:hpredict}
    \hpredict{n,j}\!:=\mathbb{E}\big( \htrue{n,j} \big) =
        \dictionary{j}(\stateEstimate{n}) \, 
        \alphavecHat{n,j} % | \stateEstimate{n} 
\quad \in \complexset{\Nantennas}{1}
    \, .
\end{align}

{\slshape Fused CSI.} The observed vector $\hmeas{n,j}$ of estimated \gls{csi} is described through the likelihood $p(\hmeas{n,j}|\htrue{n,j},\sigma_{\scriptscriptstyle j}^2)\!=\!\mathcal{CN}(\htrue{n,j}, \sigma_{\scriptscriptstyle j}^2 \eye{\Nantennas})$, i.e., corresponding to \textit{true} \gls{csi} $\htrue{n,j}$ observed in spatially uncorrelated circular \gls{awgn} with per-antenna noise variance $\sigma_{\scriptscriptstyle j}^2$.
Our predicted \gls{csi} $\hpredict{n,j}$ enters as the mean of the prior \gls{pdf} $p(\htrue{n,j}|\state{n},\alphavec{n,j})= \mathcal{CN}(\dictionary{j}(\state{n})\mualpha{n,j},\dictionary{j}\sourceCov{n,j}\dictionary{j}^\herm)$ after concentration.
Now we seek the data fusion, i.e., the fused \gls{csi} $\hfused{n,j}:= \mathbb{E}\big( \htrue{n,j} | \hmeas{n,j} \big)$, computed from the posterior \gls{pdf} (where conditional independences from Fig.\,\ref{fig:factor-graph} hold) 
\begin{align}
%\begin{split}
    p\big(&\htrue{n,j} | \hmeas{n,j}, \state{n}, \alphavec{n,j}, \sigma_{\scriptscriptstyle j}^2 \big) = 
    \frac{p\big(\htrue{n,j}, \hmeas{n,j} | \state{n}, \alphavec{n,j}, \sigma_{\scriptscriptstyle j}^2\big)}%
    {p\big(\hmeas{n,j} | \state{n},\alphavec{n,j},\sigma_{\scriptscriptstyle j}^2\big) } 
    \nonumber
    \\
    & \propto 
    p\big(\hmeas{n,j} | \htrue{n,j}, \sigma_{\scriptscriptstyle j}^2\big)
    ~
    p\big(\htrue{n,j} | \state{n}, \alphavec{n,j}\big) \label{eq:channel-posterior}
%\end{split}
\end{align}
through concentration at $\stateEstimate{n}$, where our (profile) likelihood becomes $p\big(\hmeas{n,j} | \htrue{n,j}\big)\!=\!\mathcal{CN}\big(\htrue{n,j},\sigmaHat{j}\eye{\Nantennas}\big)$ and our (profile) prior \gls{pdf} becomes $p\big(\htrue{n,j} \big)\!=\!\mathcal{CN}\big(\dictionary{j}(\stateEstimate{n})\alphavecHat{n,j}, \dictionary{j}\Phat{n}{j}\dictionary{j}^\herm \big)$.
Given the product of the circular Gaussian likelihood and circular Gaussian prior \gls{pdf} in~\eqref{eq:channel-posterior}, fused \gls{csi} is found as the posterior mean~\cite[Sec.\,8.1.8]{Cookbook}
\begin{align}
    \begin{split}\label{eq:hfused}
    \hfused{n,j} &= \mathbb{E} \big( \htrue{n,j}| \hmeas{n,j} \big) \\
    &= 
    \bm{R}_{\text{\tiny f},\scriptscriptstyle n,j}
    \big( 
        \bm{R}_{\text{\tiny m},\scriptscriptstyle n,j}^{-1} \hmeas{n,j}
        +
        \bm{R}_{\text{\tiny p},\scriptscriptstyle n,j}^{-1} \hpredict{n,j} %\htrue{n,j}
    \big) 
    \end{split}
\end{align}
of $\htrue{n,j}| \hmeas{n,j}$ which is likewise circular Gaussian-distributed (up to a normalization constant) with covariance matrix $\bm{R}_{\text{\tiny f},\scriptscriptstyle n,j}\!=\!
    \big( 
        \bm{R}_{\text{\tiny m},\scriptscriptstyle n,j}^{-1} 
        +
        \bm{R}_{\text{\tiny p},\scriptscriptstyle n,j}^{-1} 
    \big)^{-1}$ 
and is found as the probabilistic data fusion of estimated \gls{csi} with covariance matrix $\bm{R}_{\text{\tiny m},\scriptscriptstyle n,j}\!=\!\sigmaHat{j} \eye{\Nantennas}$ and predicted \gls{csi} with covariance matrix $\bm{R}_{\text{\tiny p},\scriptscriptstyle n,j}\!=\!\dictionary{j} \Phat{n}{j} \dictionary{j}^\herm 
$.
Due to $\rank\big( \dictionary{j} \Phat{n}{j} \dictionary{j}^\herm \big) \leq \Ncomponents \ll \Nantennas$,~\eqref{eq:hfused} suffers from $\bm{R}_{\text{\tiny p},\scriptscriptstyle n,j}$ being singular and hence $\bm{R}_{\text{\tiny p},\scriptscriptstyle n,j}^{-1}$ does not exist.

\begin{proposition}
    The channel fusion in~\eqref{eq:hfused} corresponds to the \gls{lmmse} estimator~\cite[Theorem\,10.3]{kay1993estimation}\,\footnote{Formulated for amplitudes $\alphavec{\,\!}$, the \gls{lmmse} estimator $\mathbb{E}(\alphavec{\,}|\state{\,},\widetilde{\bm{h}})$ in~\cite{kay1993estimation}  needs to be left-multiplied by $\bm{\Psi}(\state{\,\!})$ and concentrated using $\widehat{\bm{\alpha}}|\state{\,}$, $\widehat{\bm{P}}|\state{\,}$ to result in $\mathbb{E}(\bm{h}|\bm{\theta},\widetilde{\bm{h}}) = \bm{\Psi}\widehat{\bm{\alpha}}+\bm{\Psi}\widehat{\bm{P}}\bm{\Psi}^\herm(\sigma^2\eye{}+\bm{\Psi}\widehat{\bm{P}}\bm{\Psi}^\herm)^{-1}(\widetilde{\bm{h}} - \bm{\Psi}\widehat{\bm{\alpha}})$, which is equivalent to~\eqref{eq:lmmse}, a common result in the literature~\cite{Ozdogan19MMSE,Larsson07MMSEsparse}.
    }
\end{proposition}
\vspace{-0.65cm}%
\begin{align}\label{eq:lmmse}
        \hfused{n,j} = \hpredict{n,j} + 
        \bm{R}_{\text{\tiny p},\scriptscriptstyle n,j}
        \big( \sigmaHat{j} \eye{\Nantennas} + \bm{R}_{\text{\tiny p},\scriptscriptstyle n,j} \big)^{-1}
        (\hmeas{n,j} - \hpredict{n,j})\,.
\end{align}
\begin{proof}
See Appendix~\ref{app:LMMSE-channel-estimator}.
\end{proof} %
Using~\eqref{eq:lmmse} instead of~\eqref{eq:hfused} allows the computation of $\hfused{n,j}$ even with rank-deficient $\bm{R}_{\text{\tiny p},\scriptscriptstyle n,j}$. 
Again, the inverse in~\eqref{eq:lmmse} is efficiently implemented using~\eqref{eq:invR-eff}.

\section{\MakeUppercase{Results}}\label{sec:results}
% ---------- Results ----------
\newcommand{\MCrealizations}{1000} % Adapt this with final setting
\newcommand{\hRMSE}{6.37} % Adapt this with final setting
\newcommand{\vRMSE}{4.92} % Adapt this with final setting
% ---------- Plot Settings ----------
\newcommand{\markRepeat}{15}
\newcommand{\markSize}{1}
% ---------- ---------- ----------
Given the scenario in Fig.\,\ref{fig:TrajectoryScenario}, we choose a fixed model order, i.e., a constant state-vector dimension, representing only $S\!=\!2$ planar surfaces (\glspl{mva}), i.e., $\PMVAposStacked{n} = \big[{\pmva{1}}^{\trp}\!, {\pmva{2}}^\trp\big]^{\!\trp}$\!.

\begingroup
    \renewcommand{\thesubsection}{Experiment \Alph{subsection}}%
    \renewcommand{\thesubsectiondis}{Experiment \Alph{subsection}.}%
  
\subsection{Inverse Problem} 

Our \gls{pf} is initialized by drawing $\Nparticles\!=\!1000$ particles from $\mathcal{U}(\state{\text{\tiny min}},\state{\text{\tiny max}})$ and given the observations $\{\observationMatrix{n}\}_{\scriptscriptstyle n=1}^{\scriptscriptstyle N}$.
We initialize by drawing particles uniformly from the spatial region between $[\state{\text{\tiny min}}]_{\scriptscriptstyle 1:3}\!=\![4,-4,0]^\trp \SI{}{\metre}$ and $[\state{\text{\tiny max}}]_{\scriptscriptstyle 1:3}\!=\![12,0,3]^\trp \SI{}{\metre}$, covering the chosen scenario.
We perform \num{\MCrealizations} estimation runs with random particles $\{\particle{n}{i}\}_{\scriptscriptstyle i}^{\scriptscriptstyle \Nparticles}$ and random noise $\noise{n}{j}\!\sim\!\mathcal{CN}(\bm{0},\sigma_{\scriptscriptstyle j}^2 \eye{\Nfrequency\Nantennas})$ in each trajectory step, where we set the (time-constant) noise power $\sigma_{\scriptscriptstyle j}^2\!=\!\sigma_{\scriptscriptstyle j'}^2\, \forall j,j'\!\in\!\{1\dots J\}$ to satisfy $\SNR(n\!=\!1)\!=\!\SI{-6}{\dB}$ at time $n\!=\!1$, which decreases $\SNR(n)\!<\!\SI{-6}{\dB}$ for $n\gg1$ due to the distance-dependent path loss and \gls{olos} conditions. 
Here, $\SNR$ denotes the channel (input) \gls{snr}~\cite{Deutschmann23ICC} which we define as $\SNR(n)\!=\!\frac{1}{\Nrw}\sum_{\scriptscriptstyle j = 1}^{\scriptscriptstyle J}\frac{\lVert \htrue{n,j} \rVert^2 / \Nantennas}{\sigma_{\scriptscriptstyle j}^2}$.
We only utilize $\Nfrequency\!=\!6$ subcarrier frequencies equally spaced across the bandwidth $B$ centered around $\fc$ to perform channel estimation in the inverse problem. 
Fig.\,\ref{fig:MC-results} shows the estimated trajectories $\{\posEstimate{n}\}_{\scriptscriptstyle n=1}^{\scriptscriptstyle N}$ (\customlineref{color=\trajectoryEstimateCOL, line width=2*\trajectoryEstimateLW, line join=round, line cap=round,draw opacity=0.4}) in comparison to the ground truth $\{\pos{n}\}_{\scriptscriptstyle n=1}^{\scriptscriptstyle N}$ (\customlineref{color=IEEEblue, line width=\trajectoryLW, line join=round, line cap=round}). 
Fig.\,\ref{fig:CDF} shows the cumulative frequency of the horizontal position error $\lVert [\posEstimate{n}]_{\scriptscriptstyle1:2}\!-\![\pos{n}]_{\scriptscriptstyle1:2} \rVert $ (\,\ref{pgf:cdf-horizontal}\,)
and the vertical position error $\lVert [\posEstimate{n}]_{\scriptscriptstyle3}\!-\![\pos{n}]_{\scriptscriptstyle3} \rVert$ (\,\ref{pgf:cdf-vertical}\,) over all $N$ time steps of all \num{\MCrealizations} realizations. 
After convergence, our estimator achieves an overall horizontal position \gls{rmse} of \SI{\hRMSE}{\centi\metre} (\,\ref{pgf:RMSE-horizontal}) and vertical position \gls{rmse}%
\ifthenelse{\equal{\shortPaperVersion}{true}}%
{
    % skip in short paper version
}{%
\footnote{Dominated by a possibly hardware-related bias of \SI{4}{\centi\metre}, see Fig.\,\ref{fig:CDF}.} 
}%
of \SI{\vRMSE}{\centi\metre} (\,\ref{pgf:RMSE-vertical}) using the stochastic likelihood in~\eqref{eq:lhf-sto-profile}.
Using the deterministic likelihood in~\eqref{eq:lhf-det-profile}, the \glspl{rmse} become \SI{6.54}{\centi\metre} and \SI{4.97}{\centi\metre}, respectively. 
As noted in the literature~\cite[p.\,78]{KrimViberg96ASP}, significant differences can be observed under low \gls{snr}, highly correlated signals, and small numbers of antennas, the latter of which is not the case in our scenario.
Note, however, that~\eqref{eq:lhf-det-profile} is computationally more efficient than~\eqref{eq:lhf-sto-profile}.
Using an \gls{los}-only model, the performance degrades significantly both in the horizontal (\,\ref{pgf:cdf-horizontalLoS}) and vertical (\,\ref{pgf:cdf-verticalLoS}) errors (\SI{41}{\centi\metre} and \SI{8.4}{\centi\metre} \gls{rmse}). 
\ifthenelse{\equal{\shortPaperVersion}{true}}%
{
    % skip in short paper
}{
    Note that the limited extent of wall $s\!=\!2$ imposes the following problems with our current algorithm: i) The respective \gls{mva} can only be inferred through a second-order reflection $\MVApair{}{}$.
    ii) Our model assumes an infinite extent for which we manually need to remove component $\MVApair{2}{2}$ from the dictionary $\dictionary{j}$ to ensure a robust operation.
}%

%\textit{Forward problem.} 
\subsection{Forward Problem} 
\newcommand{\csi}[1]{\widehat{\channel}_{\scriptscriptstyle #1}}                            % a general CSI
In Fig.\,\ref{fig:channel-fusion} we evaluate the efficiency, i.e., the \gls{pg}\footnote{Note that $\pathgain\!=\!P_{\text{\tiny r}}/P_{\text{\tiny t}}$ is the ratio of receive power $P_{\text{\tiny r}}$ to transmit power $P_{\text{\tiny t}}$.} 
$\pathgain\big(\{\csi{n,j}\}_{\scriptscriptstyle j=1}^{\scriptscriptstyle J} \big)\!=\!\big|\sum_{\scriptscriptstyle j=1}^{\scriptscriptstyle J} \csi{n,j}^\herm \, \htrue{n,j}/\lVert \csi{n,j} \rVert \big|^2$, of a conjugate beamformer, 
with $\csi{n,j}\!\in\!\big\{\hmeas{n,j}, \hpredict{n,j}, \hfused{n,j} \big\}$ being any of our three types of estimated, predicted, or fused \gls{csi}: 
We have shown that the expected efficiency of a reciprocity-beamformer (\,\ref{pgf:PGestimateExpected}\,) given noisy estimated \gls{csi} $\hmeas{n,j}$ incurs an efficiency loss\footnote{Actually $0 < \SNR/(1\!+\!\SNR) < 1$ is a multiplicative gain.} of $\SNR/(1\!+\!\SNR)$ w.r.t. perfect \gls{csi}~\cite[eq.\,(17)]{Deutschmann23ICC}.
The realized efficiency $\pathgain\big(\{\hmeas{n,j}\}_{\scriptscriptstyle j=1}^{J}\big)=:\pathgain^{\text{\tiny m}}_{\scriptscriptstyle\!n}$ 
depends on the actual noise realization, which is why we augment the expectation with a symmetric confidence interval $U_{98\%}$ (\ref{pgf:PGconfidence}) for which one random realization of $\pathgain^{\text{\tiny m}}_{\scriptscriptstyle\!n}$ is located within the interval $[\mathbb{E}\{\pathgain^{\text{\tiny m}}_{\scriptscriptstyle\!n}\}\!-\!U_{98\%}, \mathbb{E}\{\pathgain^{\text{\tiny m}}_{\scriptscriptstyle\!n}\}\!+\!U_{98\%}]$ with a confidence level of ${98\%}$, i.e., the probability ${98\%}\!=\!\mathbb{P}(\mathbb{E}\{\pathgain^{\text{\tiny m}}_{\scriptscriptstyle\!n}\}\!-U_{98\%}\!\leq\!\pathgain^{\text{\tiny m}}_{\scriptscriptstyle\!n}\!\leq\!\mathbb{E}\{\pathgain^{\text{\tiny m}}_{\scriptscriptstyle\!n}\}\!+\!U_{98\%})$. 
For $\SNR\!\leq\!\SI{-6}{\dB}$, the reciprocity-beamformer incurs a loss of $\geq\!\SI{6.98}{\dB}$ w.r.t. perfect \gls{csi}.
Our geometry-based beamformer given predicted \gls{csi} (\,\ref{pgf:PGpredict}\,) outperforms the reciprocity beamformer by \SI{6.5}{\dB} on average for the chosen \gls{snr}. 
Before convergence of our \gls{pf} or under a model mismatch through e.g. assuming an \gls{los}-only model under \gls{olos} conditions as shown in Fig.\,\ref{fig:channel-fusion}\,b), estimated \gls{csi} outperforms predicted \gls{csi}.
The benefits of fused \gls{csi}~(\,\ref{pgf:PGfused}\,) 
from~\eqref{eq:hfused} become evident when comparing Fig.\,\ref{fig:channel-fusion}\,a) showing the full model with two \glspl{mva} with Fig.\,\ref{fig:channel-fusion}\,b) showing the \gls{los}-only model (i.e., $\Ncomponents\!=\!1$): %.(i.e., $\Nwalls\!=\!0$): 
Fused \gls{csi} performs \textit{robustly} in resorting to estimated \gls{csi} or predicted \gls{csi} if it significantly outperforms the respective other,\!\footnote{In such regimes, fused \gls{csi} incurs a small loss w.r.t. the respective outperforming \gls{csi}.} \!e.g., during convergence in Fig.\,\ref{fig:channel-fusion}\,a).
Fused \gls{csi} gains \textit{efficiency} if both types of estimated \gls{csi} and predicted \gls{csi} show comparable performance, as can be observed in the case of the model mismatch in Fig.\,\ref{fig:channel-fusion}\,b).
Under device \textit{mobility}, \gls{csi} ages from one time step to the next which materializes as another significant loss incurred by a reciprocity-beamformer given outdated \gls{csi}~(\,\ref{pgf:PGoutdated}\,). 
This contrasts with a geometry-based beamformer given future predicted \gls{csi}, i.e., first predicting the future state using~\eqref{eq:state-space-model} and using it to predict future \gls{csi} through \eqref{eq:hpredict}, resulting in only a marginal efficiency loss~(\,\ref{pgf:PGpredict-future}\,). 

\newcommand{\useMonteCarloResults}{true}
\begin{figure}[tp]
\centering
\ifthenelse{\equal{\useMonteCarloResults}{true}}
    {
    \centering
    \includegraphics[trim=5 10 0 0,clip,width=0.99\columnwidth]{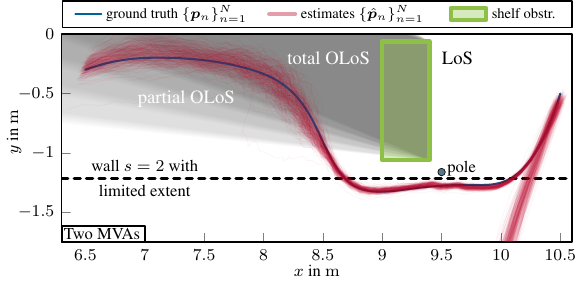}
    }{
      figure excluded
    }%
\vspace{-0.6cm}\caption{Estimated trajectories of \num{300} %\num{\MCrealizations} 
\gls{mc} estimation runs versus the true trajectory.
}\label{fig:MC-results}\vspace{-0.15cm}
\end{figure}

\begin{figure}[t]
\vspace{-0.2cm}     
\centering
\setlength{\figurewidth}{0.89\columnwidth}
\setlength{\figureheight}{0.25\columnwidth}
\input{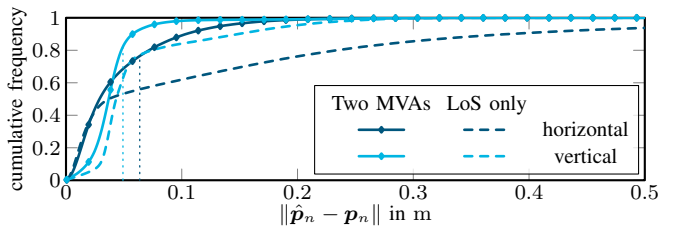}
\vspace{-0.4cm}
\caption{
Cumulative frequency of the error of \gls{mmse} estimates w.r.t. the ground truth for all time instances $n$. % along the trajectory.
}
\label{fig:CDF}
\vspace{-0.5cm}
\end{figure}

\begin{figure}[t]
\centering
\setlength{\figurewidth}{0.94\columnwidth}
\input{figures/channel-fusion-6GHz}%
\vspace{-0.5mm}%
\input{figures/channel-fusion-LoS-only}
\vspace{-0.5cm}
\caption{%Empirical \acrshort{cdf} 
Mean efficiencies ${\pathgain}$ 
of a conjugate beamformer given estimated \gls{csi} $\hmeas{n}$, predicted \gls{csi} $\hpredict{n}$, or fused \gls{csi} $\hfused{n}$.}
\label{fig:channel-fusion}
\vspace{-0.5cm}
\end{figure}

\section{\MakeUppercase{Conclusions}}\label{sec:conclusions}
In the inverse problem, we used a direct \gls{slam} approach to infer the agent position and environment map on the uplink.
In the forward problem, we used the inferred parameters to predict \gls{csi} on the downlink.
Answering on our hypotheses formulated in Sec.\,\ref{sec:scenario}, we draw the following conclusions:
\begin{enumerate}[leftmargin=16.5pt]
    \item[\circleblue{1}] \textit{Robustness}: We achieved both robust positioning in the inverse problem (see Fig.\,\ref{fig:CDF}), and robust beamforming in the forward problem (see Fig.\,\ref{fig:channel-fusion}\,a) by exploiting \gls{nlos} paths to bypass \gls{olos} conditions.
    \item[\circleblue{2}] \textit{Efficiency}: Indeed, predicted \gls{csi} can outperform estimated \gls{csi} under i) low \gls{snr}. 
    Fused \gls{csi} outperformed the other two \textit{only} if their performance was of a similar order of magnitude in efficiency. 
    \item[\circleblue{3}] \textit{Mobility Support}: Another practically relevant use for \gls{csi} prediction is ii) user mobility, where estimated \gls{csi} suffers significant losses due to \gls{csi} aging whereas \gls{csi} predicted to future time steps incurs only a marginal loss.
\end{enumerate}
\ifthenelse{\equal{\shortPaperVersion}{true}}%
{%      Omit in the short paper
}{
These findings highlight that a combination of geometry-based channel models with positioning and mapping can improve downlink transmission efficiency---improving data rates in a communication context and transmission efficiency in a power transfer context---and increase robustness---improving reliable connectivity even in \gls{olos} conditions and under user mobility.

Our current approach estimates amplitudes directly from the data and hence suffers from the impact of observation noise. 
We expect that a more accurate amplitude model will enable further efficiency gains, particularly in \gls{olos} conditions where the \gls{snr} is very low.
}
\endgroup

\begin{appendices}  % ---------- ---------- ----------

\section*{Appendix}
%\addcontentsline{toc}{section}{Appendices}
\renewcommand{\thesubsection}{\Alph{subsection}}

\titlespacing*{\subsection}{0pt}{1ex}{0ex}

\subsection{Unbiasedness}\label{app:unbiasedness}
We demonstrate that our \gls{ml} estimators concentrating the stochastic likelihood function in~\eqref{eq:lhf-sto} from Sec.\,\ref{sec:sto-lhf} are unbiased.
Per~\eqref{eq:observation}, using the statistical model from Sec.\,\ref{sec:sto-lhf} it is clear that
$%\begin{align}
    \mathbb{E}\left( \observation{n}{j}|\state{n} \right) = 
    \mathbb{E}\left( 
        \dictionary{j} 
        \alphavec{n,j}  
        + \noise{n}{j}
    \right)
    = \dictionary{j} 
      \mualpha{n,j}
$%\end{align}
and
\begin{align}
    %\begin{split}
    \mathbb{E}\left( \observation{n}{j}
    {\observation{n}{j}}^\herm|\state{n} \right) 
    &= 
    \mathbb{E}\big( 
    \dictionary{j}\alphavec{n,j}\alphavec{n,j}^\herm\dictionary{j}^\herm
    + \dictionary{j}\alphavec{n,j} {\noise{n}{j}}^\herm \nonumber \\
    &~~~+ {\noise{n}{j}} \alphavec{n,j}^\herm \dictionary{j}^\herm 
    + {\noise{n}{j}} {\noise{n}{j}}^\herm
    \big) \\
    &= \dictionary{j}
    \underbrace{\mathbb{E}\big(\alphavec{n,j}\alphavec{n,j}^\herm\big)}_{\triangleq \sourceCov{n,j}}
    \dictionary{j}^\herm
    + \underbrace{\mathbb{E}\big({\noise{n}{j}} {\noise{n}{j}}^\herm\big)}_{\triangleq \sigma_{\scriptscriptstyle j}^2 \eye{\Nfrequency\Nantennas}}
     \nonumber
    %\end{split}
\end{align}
satisfying $\observation{n}{j}|\state{n}\!\sim\!\mathcal{CN}(\dictionary{j}(\state{n}) \mualpha{n,j}, \signalCov{n,j})$.
Our amplitude mean estimator in~\eqref{eq:sto-alpha-hat} is unbiased as its expectation 
\begin{align}
    \mathbb{E}\left( 
    \alphavecHat{n,j} | \state{n}
    \right) 
    = 
    \dictionary{j}^\dag
    \, \mathbb{E}\big( \observation{n}{j} \big)
    = \dictionary{j}^\dag \dictionary{j} \mualpha{n,j} = \mualpha{n,j}
\end{align}
is the amplitude mean.
Using $\PiPerp{j}\mathbb{E}\big( \observation{n}{j}{\observation{n}{j}}^\herm \big)\!=\!(\eye{}\!-\!\PiPar{j})\signalCov{n,j}\!=\!\sigma_{\scriptscriptstyle j}^2\eye{}\!-\!\sigma_{\scriptscriptstyle j}^2\PiPar{j}$, with the parallel projector $\PiPar{j}=\dictionary{j}\dictionary{j}^\dag$, the noise variance estimator in~\eqref{eq:sto-sigma2-hat} is unbiased as its expectation is
\begin{align}
    \mathbb{E}\left( 
    \sigmaHat{j} | \state{n}
    \right) 
    &= \frac{1}{\Nfrequency \Nantennas\!-\!\Ncomponents} \tr \left(
    \sigma_{\scriptscriptstyle j}^2\eye{\Nfrequency\Nantennas}\!-\!\sigma_{\scriptscriptstyle j}^2\dictionary{j}\dictionary{j}^\dag
    \right) 
    \nonumber \\
    &= \frac{1}{\Nfrequency \Nantennas\!-\!\Ncomponents}  \left(
    \tr\big(\sigma_{\scriptscriptstyle j}^2 \eye{\Nfrequency\Nantennas}\big)\!-\tr\big(\!\sigma_{\scriptscriptstyle j}^2\dictionary{j}^\dag\dictionary{j}\big) 
    \right) 
    \nonumber \\
    &= \frac{1}{\Nfrequency \Nantennas\!-\!\Ncomponents}  \left(\sigma_{\scriptscriptstyle j}^2
    \tr\big( \eye{\Nfrequency\Nantennas}\big)\!-
    \sigma_{\scriptscriptstyle j}^2
    \tr\big(\!\eye{\Ncomponents}\big) 
    \right)
    \nonumber \\
    &= \frac{\sigma_{\scriptscriptstyle j}^2\Nfrequency \Nantennas\!-\!\sigma_{\scriptscriptstyle j}^2\Ncomponents}{\Nfrequency \Nantennas\!-\!\Ncomponents} = \sigma_{\scriptscriptstyle j}^2 \,.
\end{align}
Using this result, our source covariance estimator in~\eqref{eq:sto-P-hat} is unbiased as its expectation is
\begin{align}
    \mathbb{E}\left( 
    \Phat{n}{j} | \state{n}
    \right) 
    &= 
    \dictionary{j}^\dag 
    \left(
        \mathbb{E}\big(\observation{n}{j}{\observation{n}{j}}^\herm\big) - \mathbb{E}\big(\sigmaHat{j} \big)\eye{\Nfrequency\Nantennas}
    \right)
    {\dictionary{j}^\dag}^\herm 
    \nonumber \\
    &= 
    \dictionary{j}^\dag 
    \left(
        \dictionary{j} \sourceCov{n,j} \dictionary{j}^\herm 
        + \sigmaHat{j} \eye{\Nfrequency\Nantennas}
        - \sigmaHat{j} \eye{\Nfrequency\Nantennas}
    \right)
    {\dictionary{j}^\dag}^\herm 
    \nonumber \\
    &= 
    \dictionary{j}^\dag 
    \dictionary{j} \sourceCov{n,j} 
    (\dictionary{j}^\dag 
    \dictionary{j})^\herm
    = \sourceCov{n,j} \,. \label{eq:sto-P-hat-efficient}
\end{align}

\subsection{LMMSE Channel Estimator}\label{app:LMMSE-channel-estimator}%
Dropping indices $\{n,j\}$ for brevity, we use the identity $(\bm{A}^{-1}+\bm{B}^{-1})^{-1} = \bm{B}(\bm{A}+\bm{B})^{-1}\bm{A}$ to manipulate~\eqref{eq:hfused} as
\begin{align}
    \hfused{\,} 
    &= 
    \big( 
        \bm{R}_{\text{\tiny m}}^{-1} 
        +
        \bm{R}_{\text{\tiny p}}^{-1} 
    \big)^{-1}\!
    \big( 
        \bm{R}_{\text{\tiny m}}^{-1} \hmeas{\,}
        +
        \bm{R}_{\text{\tiny p}}^{-1} \hpredict{\,} %\htrue{n,j}
    \big) \nonumber \\
    &= 
    \bm{R}_{\text{\tiny p}}
    \big( 
        \bm{R}_{\text{\tiny m}}
        \!+\!
        \bm{R}_{\text{\tiny p}}
    \big)^{-1}\!
    \bm{R}_{\text{\tiny m}}
    \big( 
        \bm{R}_{\text{\tiny m}}^{-1} \hmeas{\,}
        \!+\!
        \bm{R}_{\text{\tiny p}}^{-1} \hpredict{\,} %\htrue{n,j}
    \big) \nonumber \\
    &= 
    \bm{R}_{\text{\tiny p}}
    \big( 
        \bm{R}_{\text{\tiny m}}
        \!+\!
        \bm{R}_{\text{\tiny p}}
    \big)^{-1}\!
    \bm{R}_{\text{\tiny m}}
    \big( 
        \bm{R}_{\text{\tiny m}}^{-1}\hmeas{\,}
        +
        \bm{R}_{\text{\tiny m}}^{-1} (\hpredict{\,}\!-\!\hpredict{\,}) 
        +
        \bm{R}_{\text{\tiny p}}^{-1} \hpredict{\,} %\htrue{n,j}
    \big) \nonumber \\
    &= 
    \bm{R}_{\text{\tiny p}}
    \big( 
        \bm{R}_{\text{\tiny m}}
        +
        \bm{R}_{\text{\tiny p}}
    \big)^{-1}
    \big( \hmeas{\,} - \hpredict{\,} \big)
     \nonumber \\
    &~~~+
    \underbrace{\bm{R}_{\text{\tiny p}}
    \big( 
        \bm{R}_{\text{\tiny m}}
        +
        \bm{R}_{\text{\tiny p}}
    \big)^{-1}
    \bm{R}_{\text{\tiny m}}}_{=(\bm{R}_{\text{\tiny m}}^{-1} +  \bm{R}_{\text{\tiny p}}^{-1})^{-1}}
    \big(
        \bm{R}_{\text{\tiny m}}^{-1} +  \bm{R}_{\text{\tiny p}}^{-1}  %\htrue{n,j}
    \big) \hpredict{\,} \nonumber \\
    &= \hpredict{\,} + \bm{R}_{\text{\tiny p}}
    \big( 
        \bm{R}_{\text{\tiny m}}
        +
        \bm{R}_{\text{\tiny p}}
    \big)^{-1}
    \big( \hmeas{\,} - \hpredict{\,} \big)
\end{align}
showing that it is indeed equivalent to the \gls{lmmse} channel estimator in~\eqref{eq:lmmse}.

\subsection{Efficient Implementation}\label{app:implementation}

The trace term in~\eqref{eq:det-sigma2-hat} and~\eqref{eq:sto-sigma2-hat} involving a computationally expensive product of two $(\Nfrequency\Nantennas \times \Nfrequency\Nantennas)$ matrices can be simplified as
\begin{align}\label{eq:efficient-trace-term}
    \tr \left(
    \PiPerp{j} \Rhate{n}{j}
    \right) \nonumber
    &= \tr \big( (\eye{\Nfrequency\Nantennas} - \dictionary{j} \dictionary{j}^\dag) \observation{n}{j} {\observation{n}{j}}^\herm \big) \\ \nonumber
    &= \tr\big( \observation{n}{j} {\observation{n}{j}}^\herm \big) - 
    \tr\big( \dictionary{j} \dictionary{j}^\dag \observation{n}{j} {\observation{n}{j}}^\herm \big) \\ \nonumber
    &= 
    {\observation{n}{j}}^\herm \observation{n}{j} - 
    \tr \big(  \dictionary{j}^\dag \observation{n}{j} {\observation{n}{j}}^\herm \dictionary{j} \big) \\ 
    &= \lVert \observation{n}{j} \rVert^2 -
    \big( {\observation{n}{j}}^\herm \dictionary{j} \big)
    \big(  \dictionary{j}^\dag {\observation{n}{j}} \big)
\end{align}
involving two less expensive matrix-vector products.

Using Sylvester's determinant theorem~\cite[eq.\,(B.1.16)]{Pozrikidis14GridsGraphsNetworks} $\left|\eye{N}\!+\!\bm{A} \bm{B}\right| = \left|\eye{K}\!+\!\bm{B} \bm{A}\right|$ for matrices $\bm{A}\!\in\!\complexset{N}{K}$ and $\bm{B}\!\in\! \complexset{K}{N}$ as well as 
$\left|c \bm{X}\right|\!=\!c^N\left|\bm{X}\right|$ for $\bm{X} \!\in\!\complexset{N}{N}$, the computationally expensive determinant %$\left|\Rhat{n}{j}\right|$ 
of the $(\Nfrequency\Nantennas\times \Nfrequency\Nantennas)$-matrix $\Rhat{n}{j}$ in~\eqref{eq:lhf-sto-profile} can be computed as
\begin{align}\label{eq:detR-eff}
    \left|\Rhat{n}{j}\right| 
    &= 
    \Big|
        \sigmaHat{j} \eye{N} + \dictionary{j} \Phat{n}{j} \dictionary{j}^\herm
    \Big| \nonumber \\
    &= \sigmaHat{j}^{\scriptscriptstyle \Nfrequency\Nantennas}
    \Big|
        \eye{\Nfrequency\Nantennas} + \frac{1}{\sigmaHat{j}} \dictionary{j} \Phat{n}{j} \dictionary{j}^\herm
    \Big| \nonumber \\
    &= \sigmaHat{j}^{\scriptscriptstyle \Nfrequency\Nantennas} 
    \Big|
        \eye{K} + \frac{1}{\sigmaHat{j}} \Phat{n}{j} \dictionary{j}^\herm \dictionary{j} 
    \Big| 
\end{align}
reducing to the computationally more efficient determinant of a $(\Ncomponents\times \Ncomponents)$-matrix. 
Further, leveraging the Woodbury identity (inversion lemma), we likewise reduce the costly inversion 
\begin{align}\label{eq:invR-eff}
    \signalCov{n,j}^{-1} &= \Big( \sigmaHat{j}
        \eye{\Nfrequency\Nantennas} +  \dictionary{j} \Phat{n}{j} \dictionary{j}^\herm
        \Big)^{\!\!-1}\nonumber \\
        %{\color{red} \text{complete me!}} 
        &= \frac{1}{\sigmaHat{j}}\eye{\Nfrequency\Nantennas} 
        -
        \frac{1}{\widehat{\sigma_{\scriptscriptstyle j}^{\scriptscriptstyle4}}}
        \dictionary{j}
        \Big(
            \Phat{n}{j}^{-1} + \frac{1}{\sigmaHat{j}}\dictionary{j}^\herm \dictionary{j}
        \Big)^{\!\!-1}\!\dictionary{j}^\herm
\end{align}
to two $(\Ncomponents\times \Ncomponents)$-matrix inversions.

\end{appendices}    % ---------- ---------- ----------

%\section*{References}
\renewcommand{\baselinestretch}{0.96}\small\normalsize % 1.485
\bibliographystyle{IEEEtran}
\balance
\bibliography{IEEEabrv,bibliography}

\vfill\pagebreak

\end{document}